\documentclass{emulateapj}
\usepackage{graphicx}

\slugcomment{Internal document}
\shorttitle{Sub-mm properties of color-selected galaxies}
\shortauthors{Decarli et al.}

\def\Lsun{L$_\odot$}					 
\def\Msun{M$_\odot$}

\def\kms{km s$^{-1}$}
\def\lsim{\mathrel{\rlap{\lower 3pt \hbox{$\sim$}} \raise 2.0pt \hbox{$<$}}}
\def\gsim{\mathrel{\rlap{\lower 3pt \hbox{$\sim$}} \raise 2.0pt \hbox{$>$}}}

\begin{document}

\title{
An ALMA survey of sub-millimeter galaxies in the Extended {\em Chandra} Deep Field South: Sub-millimeter properties of color-selected galaxies}

\author{
R. Decarli\altaffilmark{1}, 
I. Smail\altaffilmark{2}, 
F. Walter\altaffilmark{1},
A.M. Swinbank\altaffilmark{2}, 
S. Chapman\altaffilmark{3,4},
K.E.K. Coppin\altaffilmark{5},
P. Cox\altaffilmark{6,7}, 
H. Dannerbauer\altaffilmark{8}, 
T.R. Greve\altaffilmark{9}, 
J.A. Hodge\altaffilmark{1}, 
R. Ivison\altaffilmark{10},
A. Karim\altaffilmark{2,11}, 
K.K. Knudsen\altaffilmark{12}, 
L. Lindroos\altaffilmark{12}, 
H.-W. Rix\altaffilmark{1}, 
E. Schinnerer\altaffilmark{1}, 
J.M. Simpson\altaffilmark{2}, 
P. van der Werf\altaffilmark{13}, 
A. Wei\ss{}\altaffilmark{14}}
\altaffiltext{1}{Max-Planck Institut f\"{u}r Astronomie, K\"{o}nigstuhl 17, D-69117, Heidelberg, Germany.}
\altaffiltext{2}{Institute for Computational Cosmology, Durham University, South Road, Durham, DH1 3LE, UK.}
\altaffiltext{3}{Institute of Astronomy, University of Cambridge, Madingley Road, Cambridge CB3 0HA, UK.}
\altaffiltext{4}{Department of Physics and Atmospheric Science, Dalhousie University, Coburg Road Halifax, B3H 4R2, UK.}
\altaffiltext{5}{Centre for Astrophysics, Science \& Technology Research Institute, University of Hertfordshire, Hatfield AL10 9AB, UK.}
\altaffiltext{6}{IRAM, 300 rue de la piscine, F-38406 Saint-Martin d'H\`eres, France.}
\altaffiltext{7}{Atacama Large Millimeter Array, Chile.}
\altaffiltext{8}{Universit\"{a}t Wien, Institut f\"{u}r Astrophysik, T\"{u}renschanzstrasse 17, 1180, Wien, Austria.}
\altaffiltext{9}{University College London, Department of Physics \& Astronomy, Gower Street, London, WC1E 6BT, UK.}
\altaffiltext{10}{Institute for Astronomy, University of Edinburgh, Blackford Hill, Edinburgh EH9 3HJ, UK.}
\altaffiltext{11}{Argelander-Institute of Astronomy, Bonn University, Auf dem H\"{u}gel 71, D-53121, Bonn, Germany.}
\altaffiltext{12}{Department of Earth and Space Sciences, Chalmers University of Technology, Onsala Space Observatory, Sweden.}
\altaffiltext{13}{Leiden Observatory, Leiden University, PO Box 9513, 2300, RA Leiden, Netherlands.}
\altaffiltext{14}{Max-Planck Institut f\"{u}r Radioastronomie, Auf dem H\"{u}gel 69, D-53121, Bonn, Germany.}

\email{decarli@mpia.de}

\begin{abstract}
We study the sub-mm properties of color--selected galaxies via a stacking analysis applied for the first time to interferometric data at sub-mm wavelengths. We base our study on 344 GHz ALMA continuum observations of $\sim20''$--wide fields centered on 86 sub-mm sources detected in the LABOCA Extended {\em Chandra} Deep Field South Sub-mm Survey (LESS). We select various classes of galaxies ($K$-selected, star-forming sBzK galaxies, extremely red objects and distant red galaxies) according to their optical/NIR fluxes. We find clear, $>10$-$\sigma$ detections in the stacked images of all these galaxy classes. We include in our stacking analysis {\em Herschel}/SPIRE data to constrain the dust SED of these galaxies. We find that their dust emission is well described by a modified black body with $T_{\rm dust}\approx30$ K and $\beta=1.6$ and IR luminosities of $(5-11)\times 10^{11}$ \Lsun{}, or implied star formation rates of 75--140 \Msun{} yr$^{-1}$. We compare our results with those of previous studies based on single-dish observations at 870\,$\mu$m and find that our flux densities are a factor 2--3 higher than previous estimates. The discrepancy is observed also after removing sources individually detected in ALESS maps. We report a similar discrepancy by repeating our analysis on 1.4\,GHz observations of the whole ECDFS. Hence we find tentative evidence that galaxies that are associated in projected and redshift space with sub-mm bright sources are brighter than the average population. Finally, we put our findings in the context of the cosmic star formation rate density as a function of redshift.
\end{abstract}

\section{Introduction}

A variety of tracers are used to probe star formation in distant galaxies, based on rest-frame UV luminosities, optical colors, recombination line luminosities, PAH features, dust luminosity, and radio emission \citep[e.g.,][]{condon92,kennicutt98,kewley01,yun02,bell03, brinchmann04,daddi07,salim07,sargent10,karim11,leroy12,murphy12}.
The dust continuum luminosity is of particular interest in the study of star formation in high-$z$ galaxies: as the dust spectral energy distribution (SED) shifts to higher and higher redshifts, we observe closer and closer to the peak of the dust emissivity at sub-mm wavelengths. This negative $k$-correction is such that it roughly compensates the flux dimming due to the increased luminosity distance, so that a galaxy with a fixed IR luminosity will show about the same sub-mm flux density at any redshift $1<z<6$ \citep[e.g.,][]{blain02}.

\begin{figure*}[t]
\includegraphics[width=0.99\columnwidth]{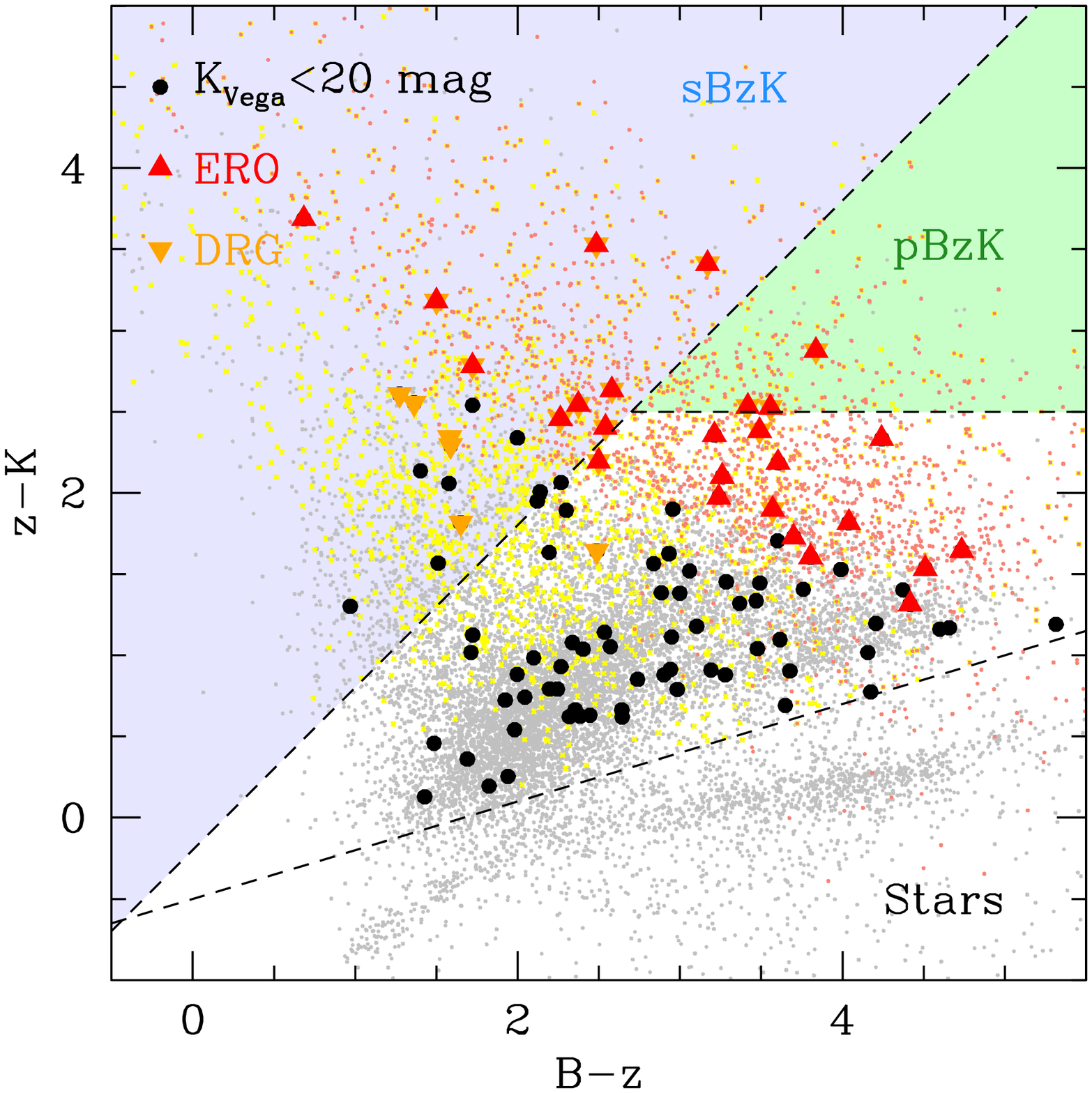}
\includegraphics[width=0.99\columnwidth]{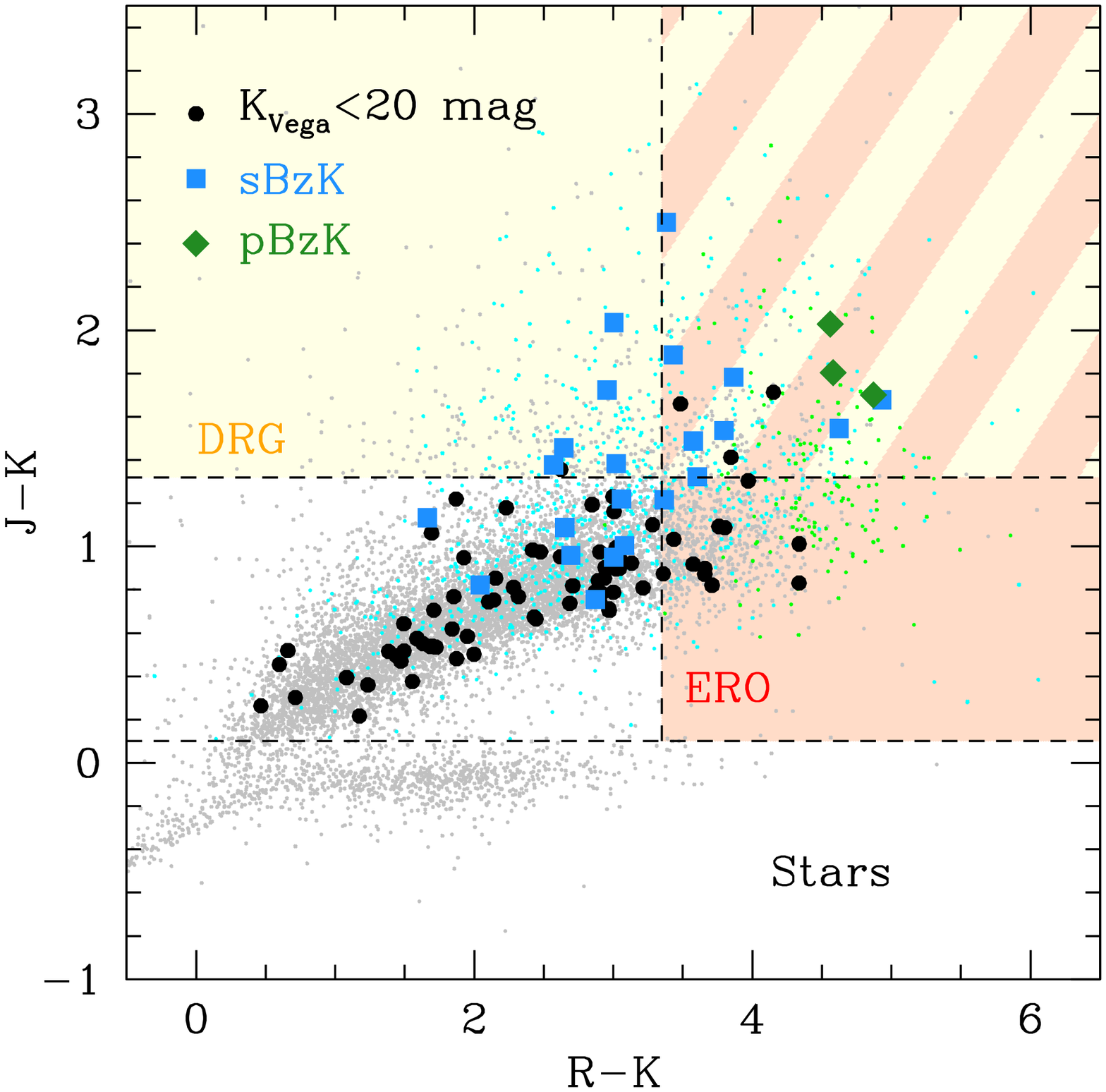}
\caption{The optical/NIR color-color diagrams used to define the different stacking samples of galaxies. Dots are sources in the photometric catalog. Circles, triangles, squares, and diamonds highlight sources which lie within the primary beam of ALMA pointings ($\sim20''$ in diameter) used in this study. {\em Left}: The BzK diagnostic allows us to identify stars from galaxies, and to isolate $1.4 \lsim z \lsim 2.5$ star-forming and passive galaxies (top left and top right corners, respectively). EROs and DRGs are highlighted with red and yellow points (upwards / downwards triangles), respectively. {\em Right}: The ($R$-$K$,$J$-$K$) color diagram allows us to distinguish stars from galaxies, and is used to define EROs and DRGs. 
Blue and green symbols (squares / diamonds) show how sBzK and pBzK galaxies distribute over this plot.}
\label{fig_coldiag}
\end{figure*}

\begin{figure*}
\includegraphics[width=0.99\columnwidth]{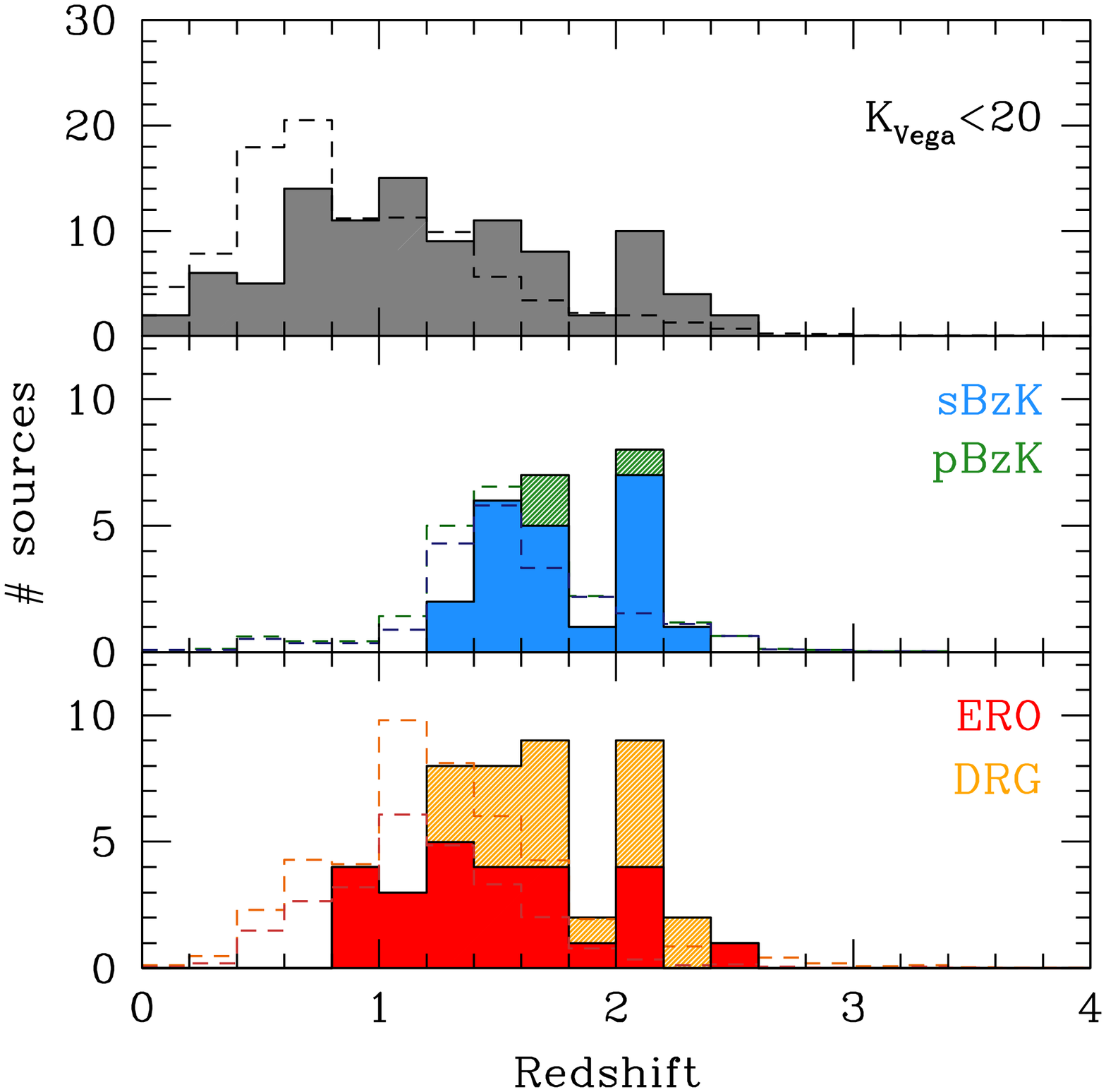}
\includegraphics[width=0.99\columnwidth]{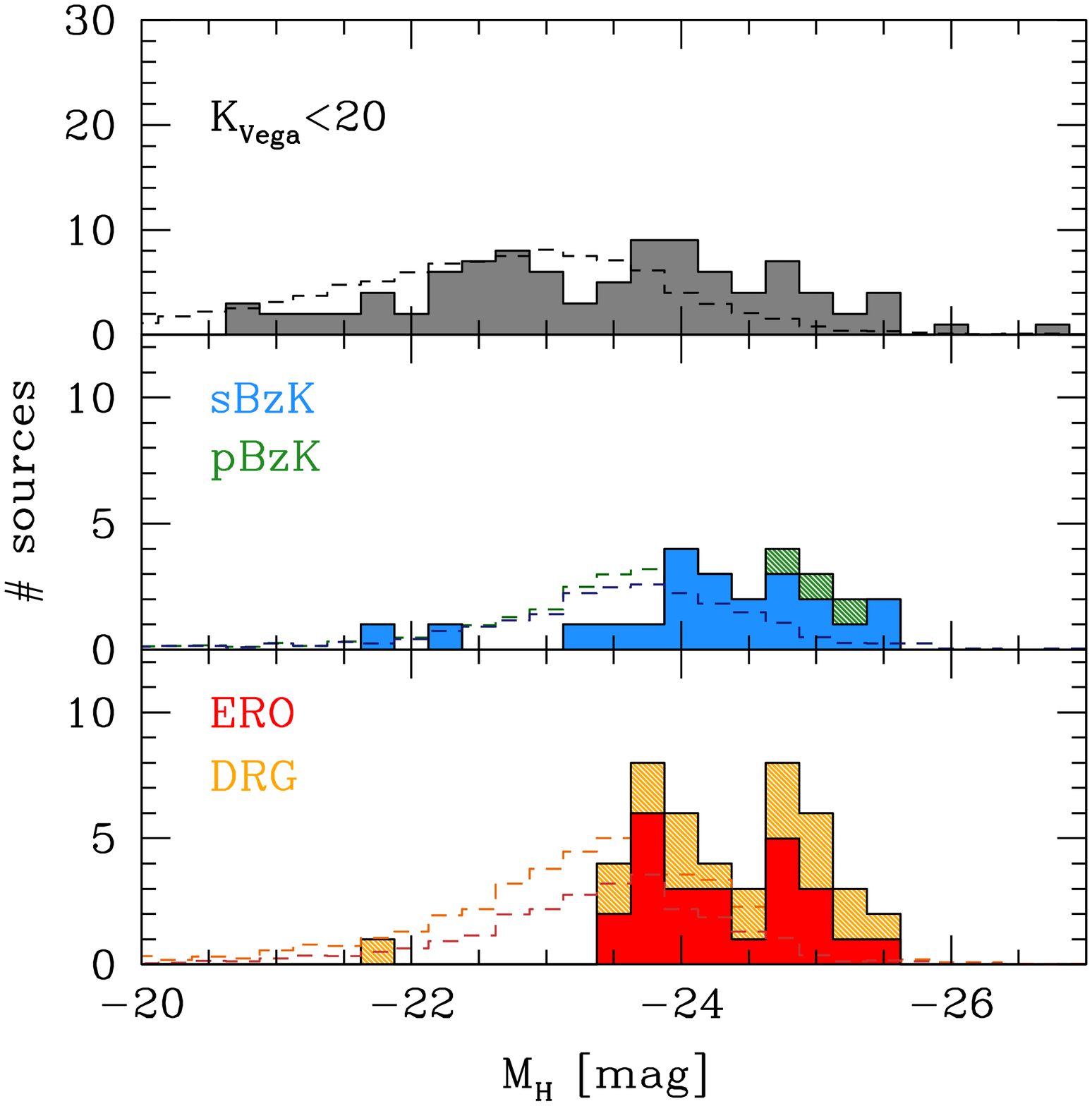}
\caption{Redshift and absolute magnitude distributions of the various samples of galaxies considered in this study. Shaded histograms show the sources used in our stacking analysis, while the dashed histograms show the parent distributions from the whole MUSYC catalog, scaled in order to match the number of galaxies (for each class) covered by ALESS pointings. {\em Left:} The general population of $K_{\rm Vega}<20$ mag galaxies is dominated by low-$z$ ($z<1.2$) sources. BzK color cuts effectively select galaxies at $1< z <2.5$, while EROs and DRGs show slightly lower redshifts on average ($0.8<z<2.4$). A small excess at $z\approx2$ is reported in all the distributions of color-selected galaxies in ALESS pointings with respect to the general field. {\em Right:} The absolute magnitude in the rest-frame $H$ band is computed via SED fitting \citep{simpson14}. The distributions observed inside and outside the ALESS coverage are similar for $K_{\rm Vega}<20$ mag galaxies, while color-selected galaxies in our analysis tend to be brighter in rest-frame $H$ band than their parent samples from the general photometric catalog.}
\label{fig_z_distr}
\end{figure*}

Sub-mm observations are thus especially suited to study star formation in high-$z$ galaxies. However, with the exception of a few, strongly lensed cases \citep[e.g.][]{knudsen09,swinbank10}, so far only the very bright end of the IR luminosity function has been constrained \citep[e.g.,][]{barger99,borys03,coppin06,weiss09,austermann10}, because of sensitivity and resolution limits of single-dish observations. These bright (several mJy at 350 GHz) sources show redshift distributions peaking around $z\gsim 2$, IR luminosities exceeding $10^{12}$ \Lsun{} and associated star--formation rates (SFRs) of hundreds or thousands of solar masses per year \citep{chapman03,chapman05,solomon05,walter09,hatsukade10,wardlow11,moncelsi11,walter12,simpson14,swinbank14}. However, these sources are not representative of the more common star--forming galaxies, with star formation rates of $\lsim10$ \Msun{} yr$^{-1}$ \citep[see, e.g.,][]{dacunha12}, which dominate the cosmic star--formation density. In order to sample these sources, sensitivities of $\lsim0.1$ mJy at 350 GHz are required. These depths are expensive to achieve (in terms of observing time) even with the full ALMA (e.g., in order to obtain a 1-$\sigma$ sensitivity of 10 $\mu$Jy at 344 GHz in a continuum observation with 50 antennas, one needs $\sim3.6$ hours on source). Moreover, ALMA observations at (sub-)mm wavelengths cover only a small region on the sky (the primary beam diameter is $17.3''$ at 344 GHz). Therefore, many pointings are required in order to address cosmic variance. 

A common way to push the sensitivity of observations of a class of faint sources is through stacking analysis of galaxies selected, e.g., via their optical/NIR emission. Various studies have applied stacking techniques basically at any wavelength: $\gamma$-rays \citep[e.g.,][]{aleksic11}, X-rays \citep{chaudhary12,george12}, optical/NIR \citep{zibetti07,matsuda12,gonzales12}, MIR/FIR \citep{dole06,kurczynski10,bourne12}, sub-mm \citep{webb04,knudsen05,greve10} and radio \citep[e.g.][]{boyle07,ivison07,hodge08,hodge09,dunne09,fabello11,karim11}. In particular, \citet{greve10} (hereafter, G10) undertook a stacking analysis of the LABOCA Extended {\em Chandra} Deep Field South Sub-mm Survey \citep[LESS;][]{weiss09}, a 870 $\mu$m (344 GHz) survey of a $30'\times30'$ wide region around the Extended {\em Chandra} Deep Field South \citep[ECDFS;][]{giacconi01}, also encompassing GOODS-South and the {\em Hubble} Ultra Deep Field. They stacked 344 GHz measurements obtained with LABOCA at the positions corresponding to galaxies grouped on the basis of their optical/NIR fluxes and colors, and of their redshift. Thanks to the large areal coverage of LESS, several hundred galaxies could be stacked in each galaxy class, thus boosting the sensitivity by more than an order of magnitude, down to few tens of $\mu$Jy. The major limit of the stacking analysis in G10 is that LABOCA observations have an intrinsic resolution of $19.2''$. This implies that source blending is a major issue \citep[see][]{swinbank14}. A sophisticated deblending algorithm based on neighbor chains was applied in G10 \citep[see also][]{kurczynski10}. 

Here we build up on the analysis by G10, and perform a similar stacking analysis on new interferometric ALMA observations of the fields encompassing LESS-detected galaxies \citep[the ALESS survey:][see \S\ref{sec_ALESS}]{hodge13,karim13}. These data have been collected at the same (effective) frequency of the original LESS observations, and reach typical rms of $\approx0.4$ mJy beam$^{-1}$, i.e., a factor 3 better than LESS. Most importantly, these interferometry observations have typical beam sizes of $\sim1.6''\times1.15''$, i.e., $\sim200$ times smaller than in the LABOCA single-dish observations (in terms of beam area). This matches the typical angular size of galaxies at high-$z$, so source blending is no longer an issue at this frequency. This enormously simplifies the interpretation of the results of any stacking analysis. 

The structure of this paper is as follows: \S\ref{sec_ALESS} briefly introduces the ALESS data. \S\ref{sec_musyc} describes the galaxy catalog and the color cuts used to define the stacked samples. We explain our stacking routine in \S\ref{sec_stack}. In \S\ref{sec_results}, we present the results of our analysis, we infer IR luminosities and SFR estimates, and we compare our findings with previous works. Our conclusions are drawn in \S\ref{sec_conclusions}.
Throughout the paper we will assume a standard cosmology model with 
$H_0=70$ \kms{} Mpc$^{-1}$, $\Omega_{\rm m}=0.3$ and 
$\Omega_{\Lambda}=0.7$. All magnitudes refer to the AB photometric system
\citep{oke74}, unless specified.

\section{The ALESS data}\label{sec_ALESS}

Our analysis is based on the ALMA survey of LESS-detected sources (ALESS). These are 344 GHz observations of 122 sources. Observations were carried out during ALMA Cycle 0 (`Early Science') between October 18 and November 3, 2011, with the array in ALMA Cycle 0's compact configuration (longest baseline: 125m), mostly with 15 antennas. The typical resolution element has a FWHM $\sim1.6''\times1.15''$. At 344 GHz, the full width at half power of the ALMA primary beam is $17.3''$ (scaling as $19.9'' \times (300/\nu)$, where $\nu$ is the observing frequency in GHz). Data were reduced and cleaned down to a 3-$\sigma$ level using the Common Astronomy Software Application ({\sc CASA}). Details on the survey design, on the data reduction and the cleaning process are described in \citet{hodge13}. Final maps are $128\times128$ pixel, with a pixel scale of $0.2''$ per pixel. The typical central rms of the final maps is $\approx0.4$ mJy beam$^{-1}$. In our analysis, we focus only on the 86 `good quality' maps with rms $<0.6$ mJy beam$^{-1}$ and beam axis ratio $<$ 2 \citep[see][for details]{hodge13}.

\section{The optical/NIR dataset}\label{sec_musyc}

We base our analysis on the photometric catalog by \citet{simpson14}, who capitalized on deep, optical/NIR archival data from various surveys of the ECDFS. The bulk of the data is taken from the Wide MUlti-wavelength Survey by Yale-Chile \citep[MUSYC; ][hereafter: T09]{taylor09}. For more details on MUSYC, see \citet{gawiser06}. The MUSYC catalog consists of 16,910 $K$-band flux limited sources in the $30'\times 30'$ wide region at the center of the ECDFS. It provides optical and NIR photometry in the $UBVRIzJHK$ bands. At $K_{\rm AB}=22$ mag the catalog is 100\% complete for point-sources, and 96\% complete for sources with a scale radius of $\approx 0.5''$ ($\approx 4.2$ kpc at $z$=2). This corresponds to a stellar mass ($M_*$) completeness of $>90$\% at $z$=1.8 for $M_*>10^{11}$ \Msun{} \citep{taylor09b}. Following G10, we focus on the $K_{\rm Vega}<20$ mag (i.e., $K_{\rm AB}<21.83$ mag) galaxies, to ease the comparison with previous studies. We take into account flux aperture corrections by scaling all the fluxes according to the SExtractor-to-total flux ratio in the $K$ band as provided in T09. Additional photometric data in \citet{simpson14} include deep $J$- and $K$-band images from Zibetti et al.~(in prep.), and from the Taiwan ECDFS NIR survey \citep[TENIS,][]{hsieh12}, and {\em Spitzer}/IRAC 3.6, 4.5, 5.8, and 8.0\,$\mu$m images from the Spitzer IRAC/MUSYC Public Legacy Survey \citep[SIMPLE,][]{damen11}.

\citet{simpson14} performed SED fitting and photometric redshift estimates for all the sources in an IRAC 3.6\,$\mu$m selected catalogue of the ECDFS. Once compared to the spectroscopic redshifts, the typical accuracy is $\Delta z /(1+z)=0.011$. In the following analysis, we will refer to the best redshift estimate (spectroscopic if available) for all the sources.

The photometric catalogue is used to select galaxies from stars and to split galaxies in various subsamples, as follows:
\begin{itemize}
\item[-] Galaxies with $K_{\rm Vega}<20$ mag, separated from stars by requiring $(z-K-0.04)>0.3 \times (B-z+0.56)-0.5$ and $(J-K)>0.1$, following \citet{daddi04} and G10. 
\item[-] Actively star-forming galaxies at $1.4< z <2.5$: dubbed as sBzK, they are defined by requiring $(z-K-0.04)-(B-z+0.56)>-0.2$ \citep{daddi04}.
\item[-] Extremely Red Objects (EROs), defined as galaxies with $(R-K)>3.35$ and $(J-K)>0.1$ \citep{elston88}.
\item[-] Distant Red Galaxies (DRGs), defined as $(J-K)>1.32$ \citep{franx03}.
\end{itemize}
All the galaxies selected as sBzK, EROs or DRGs are also part of the $K_{\rm Vega}<20$ mag sample.

\begin{figure*}
\includegraphics[width=0.99\textwidth]{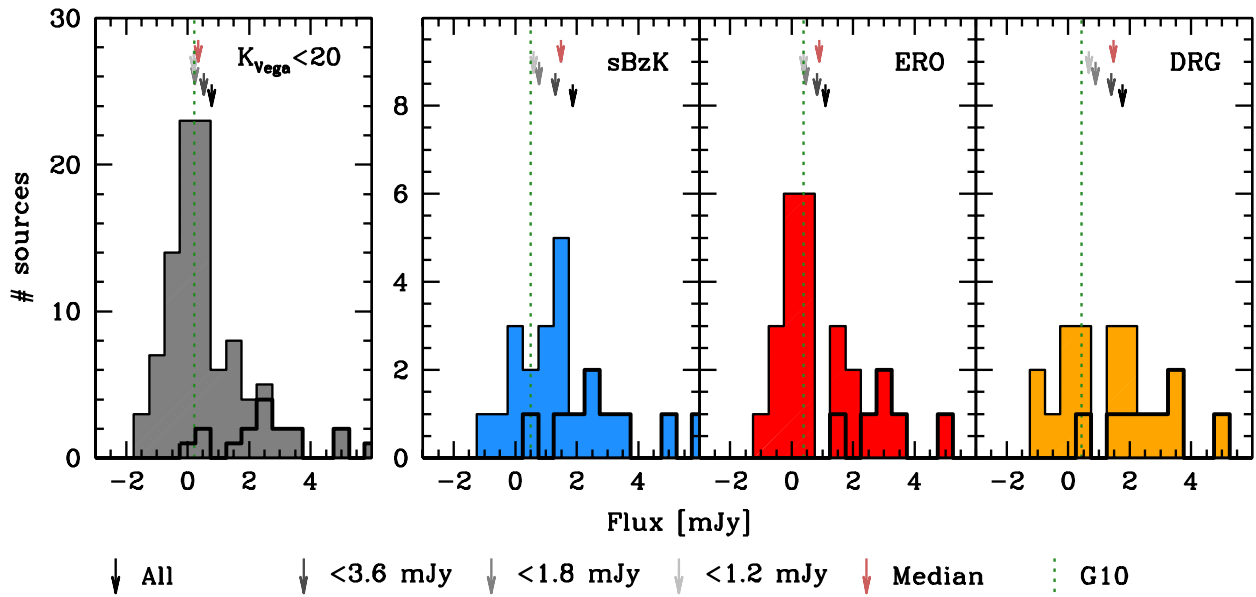}
\caption{Flux distribution (at 344\,GHz) of individual galaxies considered in this study (shaded histograms). We plot here the flux density value as measured at the positions given in the optical/NIR catalog, after correcting for primary beam attenuation. Thick lines highlight the flux distributions of galaxies physically associated with SMGs (i.e., with consistent photometric redshift and $<$200 kpc separation on sky). The flux measured from the stacking is shown as arrows, in different colors for different sub-samples (see labels and text for details). For a comparison we plot the fluxes obtained in the stacking analysis of LABOCA sources in the full ECDFS presented in \citet{greve10} as dashed lines.}
\label{fig_histo}
\end{figure*}

Fig.~\ref{fig_coldiag} shows key color cuts adopted in the definitions of our galaxy samples. The line defining sBzK galaxies runs parallel with the reddening vector, so that galaxies move up and right-wards at increasing reddening in the left-hand panel of Fig.~\ref{fig_coldiag}. This selects massive, star-forming galaxies at $1.4<z<2.5$ irrespective of dust extinction (as long as they pass the $K$-band flux selection). In the same plot, a $z>1.4$ galaxy with fixed reddening would move up and right-wards as its stellar population grows older. Therefore, the wedge defined by requiring $(z-K-0.04)-(B-z+0.56)<-0.2$ and $(z-K-0.04)>2.5$ is populated by passive, old galaxies, dubbed passive BzK galaxies \citep[pBzK;][]{daddi04}. On the other hand, the NIR color cuts used to identify EROs and DRGs select objects with very red colors, either intrinsic (i.e., associated with passive, old stellar populations at high-$z$) or due to reddening \citep{daddi04,lane07,greve10}, therefore these galaxies include high-$z$ sources with old stellar population (including pBzK galaxies) or highly--obscured star--forming galaxies. 

There is clear overlap in the sample definitions. In particular, pBzK galaxies are a subsample of the ERO class; about 1/3 of the sBzK selected galaxies are also DRGs or EROs, the fraction depending on redshift and NIR flux (most of the sBzK galaxies classified also as EROs or DRGs reside at $z>2$ and are faint in the $K$ band); about 3/4 of the DRGs are also selected as EROs. In the remainder of our analysis, we will not include pBzK galaxies, as the total number of pBzK covered in the ALESS observations is only 3.

Fig.~\ref{fig_z_distr} shows the redshift and NIR luminosity distributions of the galaxies in each sample. NIR luminosities are computed by interpolating the best SED fits from \citet{simpson14}. $K$-band selected galaxies span a wide range in redshift, from 0 to 2.6, while color selection efficiently identify sources with $1<z<2.6$. \citet{lane07} suggested that DRGs can have a broader redshift distribution than EROs, but this is not observed in our sample. Similarly, $K$-selected galaxies show a broad range of rest-frame $H$-band luminosities (from -20 to -26 mag), while color-selected galaxies tend to be bright ($M_{\rm H}<-23$ mag).

\section{The stacking routine}\label{sec_stack}

\begin{table}
\begin{center}
\caption{Summary of the stacking results.}\label{tab_results}
\begin{tabular}{ccccc}
\hline
Sample & N.gal & Flux\footnote{Stacked 344\,GHz flux, measured as the maximum flux in a pixel within $3''$ from the central pixel.} & rms\footnote{Pixel rms of the stacked image.} & Error\footnote{Stack uncertainties as estimated by stacking random coordinates in each map (see \S\ref{sec_random} for details).}   \\
       &       &  [mJy]        & [mJy]     & [mJy] \\
 (1)   & (2)   &  (3)          & (4)       & (5)   \\
\hline
\multicolumn{5}{c}{\it All}\\
$K_{\rm Vega}<20$ & 100 & 0.78 & 0.06 & 0.06 \\
sBzK	          &  22 & 1.88 & 0.15 & 0.11 \\
ERO	          &  26 & 1.11 & 0.09 & 0.10 \\
DRG	          &  20 & 1.77 & 0.13 & 0.12 \\
\hline    	 
\multicolumn{5}{c}{$S_{\nu}(344 {\rm GHz})<3.6$ mJy}\\
$K_{\rm Vega}<20$ &  97 & 0.53 & 0.06 & 0.05 \\   
sBzK	          &  20 & 1.31 & 0.14 & 0.11 \\   
ERO	          &  25 & 0.82 & 0.09 & 0.12 \\   
DRG	          &  19 & 1.41 & 0.12 & 0.11 \\   
\hline	 
\multicolumn{5}{c}{$S_{\nu}(344 {\rm GHz})<1.8$ mJy}\\
$K_{\rm Vega}<20$ &  90 & 0.23 & 0.05 & 0.05 \\   
sBzK	          &  16 & 0.77 & 0.14 & 0.14 \\   
ERO	          &  22 & 0.45 & 0.09 & 0.13 \\   
DRG	          &  15 & 0.89 & 0.13 & 0.14 \\   
\hline        
\multicolumn{5}{c}{$S_{\nu}(344 {\rm GHz})<1.2$ mJy}\\
$K_{\rm Vega}<20$ &  85 & 0.20 & 0.06 & 0.08 \\    
sBzK	          &  14 & 0.60 & 0.15 & 0.17 \\    
ERO	          &  20 & 0.39 & 0.09 & 0.13 \\    
DRG	          &  13 & 0.68 & 0.11 & 0.15 \\    
\hline
\multicolumn{5}{c}{$z>1$}\\
$K_{\rm Vega}<20$ &  52 & 1.16 & 0.09 & 0.09  \\
sBzK	          &  22 & 1.89 & 0.15 & 0.10  \\
ERO	          &  25 & 1.15 & 0.09 & 0.11  \\
DRG	          &  19 & 1.90 & 0.13 & 0.13  \\
\hline	 
\multicolumn{5}{c}{\it All, median stack}\\
$K_{\rm Vega}<20$ & 100 & 0.34 & 0.07 & 0.06 \\
sBzK	          &  22 & 1.48 & 0.15 & 0.13 \\
ERO	          &  26 & 0.89 & 0.11 & 0.10 \\
DRG	          &  20 & 1.47 & 0.14 & 0.13 \\
\hline
\multicolumn{5}{c}{\it All, stack at uniform spatial resolution}\\
$K_{\rm Vega}<20$ & 100 & 0.79 & 0.06 & 0.06 \\
sBzK	          &  22 & 1.76 & 0.15 & 0.11 \\
ERO	          &  26 & 1.02 & 0.06 & 0.10 \\
DRG	          &  20 & 1.55 & 0.11 & 0.12 \\
\hline	
\noalign{\medskip}
\hline
\multicolumn{5}{c}{Greve et al. (2010), sub-mm faint only}\\
$K_{\rm Vega}<20$ & 8209 & 0.18 & & 0.01       \\	   
sBzK	          &  725 & 0.37 & & 0.04       \\	   
ERO	          & 1228 & 0.29 & & 0.03       \\	  
DRG	          &  720 & 0.32 & & 0.04       \\	  
\hline
\multicolumn{5}{c}{Greve et al. (2010), all}\\
$K_{\rm Vega}<20$ & 8266 & 0.22 & & 0.01       \\    
sBzK	          &  744 & 0.50 & & 0.04       \\    
ERO	          & 1253 & 0.39 & & 0.03       \\	  
DRG	          &  737 & 0.43 & & 0.04       \\	  
\hline
\vspace{-0.8mm}\\
\end{tabular}
\end{center}
\end{table}

\subsection{Method}
  
\begin{figure*}
\begin{center}
\includegraphics[width=0.99\textwidth]{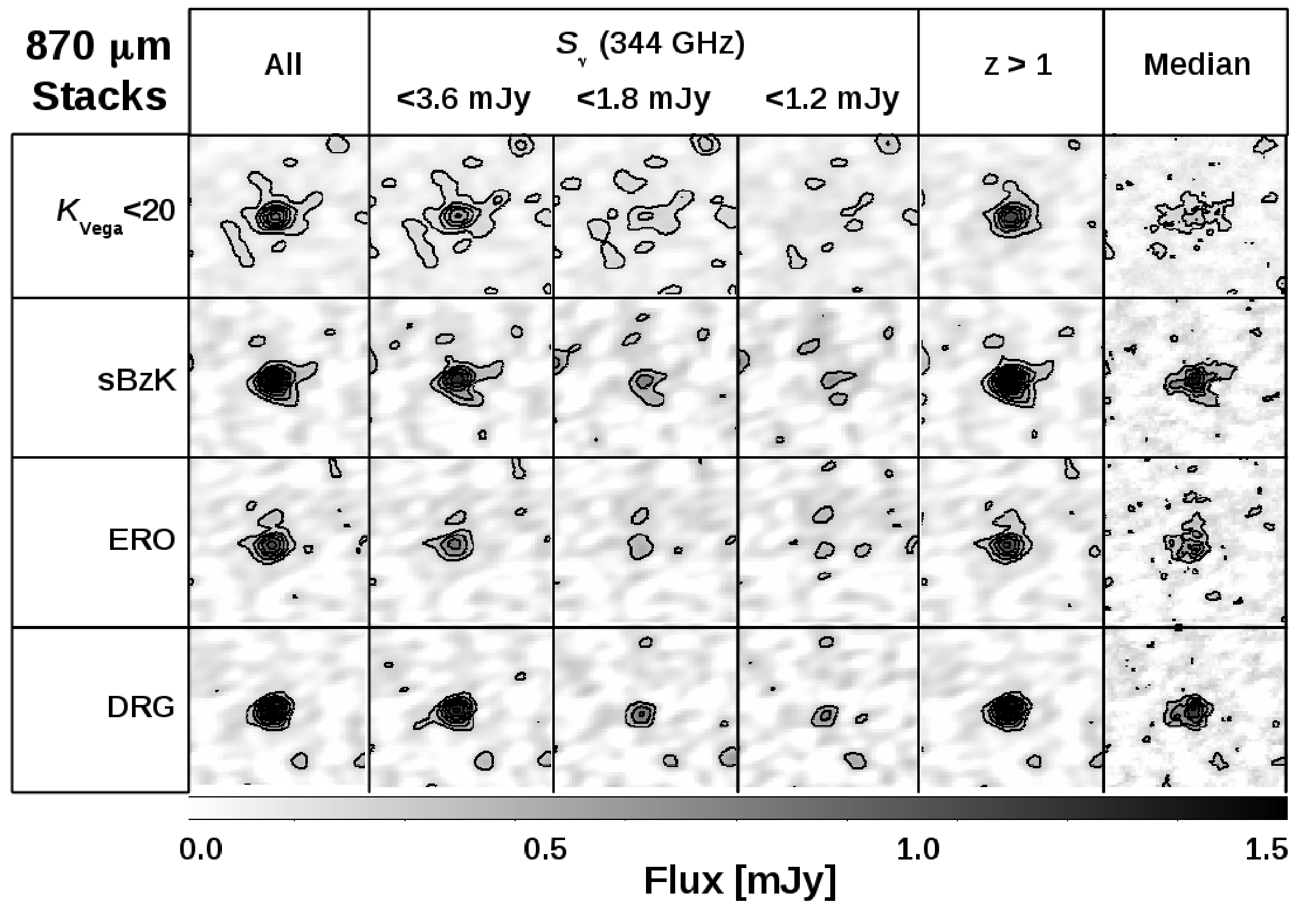}
\end{center}
\caption{Postage stamps of the stacks for all the galaxy samples considered in this study. The first column refers to the weighted-average stacks of the whole samples, while the second, third and fourth columns show the stack results of the samples after removing sources with $S_\nu$(344\,GHz)$>$3.6 mJy, 1.8 mJy and 1.2 mJy, respectively. The last two columns show the stack results for all the galaxies at $z>1$, and for the whole samples again, when median stacking is adopted instead of weighted averages. The gray scale is fixed in all the panels, while contours are 2-$\sigma$ spaced and mark the significance of the detections. Each postage stamp covers an area of $10''\times10''$. Clear detections are reported in most of the cases.}
\label{fig_stack}
\end{figure*}

Our stacking routine works as follows: 1) From the photometric catalog, we select galaxies according to their fluxes and colors (as described in \S\ref{sec_musyc}). 
We keep only sources lying within $1.2\times$ the primary beam radius of each pointing (i.e., within $10.4''$ from each pointing center), i.e., where the sensitivity is $>$1/3 of that at the pointing center. Seventy-four out of 86 `good quality' ALESS pointings overlap with the area covered by the MUSYC catalog, and 55 pointings contain at least one galaxy belonging to one of the classes defined in \S\ref{sec_musyc}. 2) We compute the offset of each galaxy with respect to the pointing center of the ALMA observations. This distance is used to compute the primary beam correction, modeled as exp(offset$^2/(2 \sigma_{PB}^2)$), where $\sigma_{PB}=PB/(2 \sqrt{2 \ln 2})$, and $PB$ is the primary beam size. 3) For each source, we create a postage stamp from the ALMA 344\,GHz image. 4) We align all the postage stamps, scale them to account for the primary beam attenuation computed at step \#2, and then weight-average them. Weights are computed as the squared inverse of the primary beam correction computed at the center of each postage stamp\footnote{Our working assumption is that all the maps have the same depth, which is true to within $\sim20$ \% accuracy, given that we consider only the `good quality' maps \citep[see Fig.~1 in][]{hodge13}.}, so that a source lying at the primary beam radius has a primary beam correction of 2 and a weight of 0.25.

We perform our stacking analysis first on all the galaxies in each bin; then, in order to attempt to account for biases in the sample selection, we exclude those sources that have 344 GHz fluxes brighter than $S_{\nu}$(344\,GHz)=$3.6$ mJy in our ALMA observations, i.e., sources that were detected at $>3$-$\sigma$ in the original LESS observations. We re-perform the stacks on this ``sub-mm faint'' sample. We then progressively lower the flux cutoff to $S_{\nu}$(344\,GHz)=$1.8$ mJy and $S_{\nu}$(344\,GHz)=$1.2$ mJy, and repeat the experiment. The first of these lower flux thresholds tentatively mimics the case where a LESS 3-$\sigma$ detection resolved into two sources in our interferometric observations\footnote{In this simplified approach, we assume equal flux splitting between the two sources.}, consistent with the finding that the bright end of the sub-mm galaxy luminosity function observed in the original LESS data is dominated by pairs or multiplets of galaxies which are unresolved in single-dish observations \citep{karim13,hodge13}. The second flux cut corresponds to a $\approx 3$-$\sigma$ detection in all the ALMA maps considered here, i.e., this subsample excludes any individually-detected ALESS source as well. Finally, we consider again the whole dataset (with no $S_\nu$(344\,GHz) cutoff), and perform 1) stacking of only those sources lying at $z>1$; and 2) median stacking instead of weighted averages of all the sources. These different tests allow us to quantify the role of outliers in our final stacks. The analysis is repeated with two different stacking routines, developed independently within our collaboration, finding consistent results. We have also tested the effects of beam variations among different ALESS pointings by artificially lowering the spatial resolution of all the maps to a circular resolution element with FWHM=$1.6''$. The results of this test are in agreement with those obtained when stacking maps in their original resolution (see Table \ref{tab_results}).

The distributions of 344 GHz fluxes of the galaxies considered in this study, as measured on the ALESS data, and the results from the stacks are shown in Fig.~\ref{fig_histo}. All the samples show a bell-shaped flux distribution with a tail towards positive fluxes, implying significant detections from the stacks of each sample (even after excluding the brightest sources).

\subsection{Estimate of uncertainties}\label{sec_random}

In order to quantify the uncertainties and biases in the results of our stacking analysis, we repeat our analysis at random positions uniformly distributed over the same area. The assumption here is that the maps are mostly ``source-free'', so that stacking random positions in the sky corresponds to a random sampling of the noise properties in the maps. For each galaxy set of $N$ sources, we create $N$ random coordinates. The distribution of sources over the various pointings is the same as for the original galaxies (e.g., if a pointing contains three sBzK, we take three random coordinates from that pointing). We stack the ALESS images of these $N$ random coordinates following the same procedure as in the case of `real' sources. Then we repeat the whole procedure with a new realization of $N$ random coordinates. We perform 50 iterations, in order to gauge the variance of the random stacks. The rms of the distribution of the final stacked fluxes is taken as a measure of the noise in the maps\footnote{If real sources were always close to the pointing center, the approach outlined here would underestimate the signal-to-noise of the stacking results, due to the primary beam correction. This effect can be estimated a posteriori (by measuring the average attenuation due to primary beam tapering in the samples of real sources, and in the case of uniform distributions). The signal-to-noise ratio is underestimated by $3-7$ \% in the various subsamples, i.e., the correction is negligible for the degree of accuracy required in our analysis.}.
We obtain the following rms values: $0.06$ mJy beam$^{-1}$,  $0.11$ mJy beam$^{-1}$, $0.10$ mJy beam$^{-1}$, $0.12$ mJy beam$^{-1}$ for $K_{\rm Vega}<20$ mag, sBzK, EROs and DRGs respectively, when weighted averages are used. Consistent values (within $\approx$10\%) are found with median stacks.
These values are also found to be in good agreement with the pixel rms computed close to the center of stacked images of real sources: $0.06$ mJy beam$^{-1}$,  $0.15$ mJy beam$^{-1}$, $0.09$ mJy beam$^{-1}$, $0.13$ mJy beam$^{-1}$ for $K_{\rm Vega}<20$ mag, sBzK, EROs and DRGs respectively. Finally, we have verified that the average stacked values in the random iterations are always consistent with zero.

\subsection{Stacks of {\em Herschel}/SPIRE observations}
\begin{figure}
\begin{center}
\includegraphics[width=0.99\columnwidth]{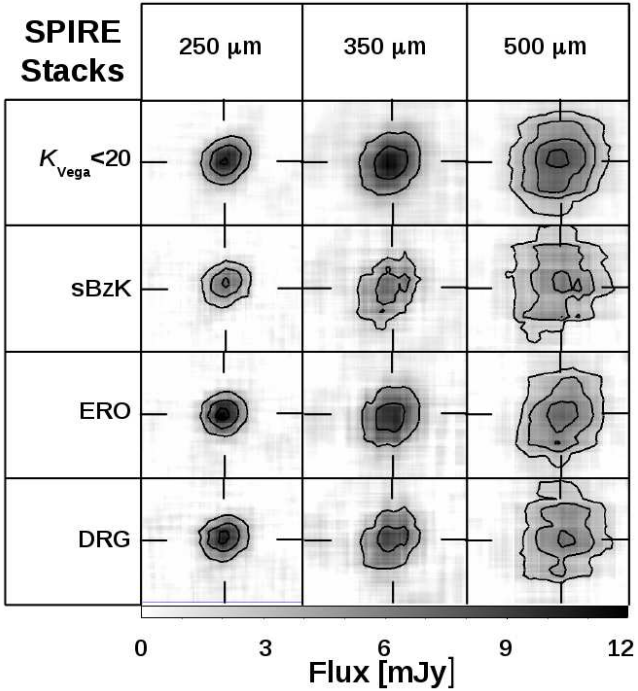}
\end{center}
\caption{Postage stamps of the stacks of {\em Herschel}/SPIRE observations of the $z>1$ samples of color--selected galaxies in our analysis. Each panel is $40''\times35''$ wide. The gray scale is fixed in all the panels, while contours are 2-$\sigma$ spaced and mark the significance of the detections. All the samples show a detection in the SPIRE bands.}
\label{fig_spire_ps}
\end{figure}

In order to calculate average IR luminosities and star formation rates, we repeat our stacking analysis on the {\em Herschel}/SPIRE maps of the ECDFS at 250, 350, and 500\,$\mu$m. These observations are part of the {\em Herschel} Multi-tiered Extragalactic Survey \citep[HerMES;][]{oliver12}. The SPIRE maps have resolution elements of $17.6''$, $23.9''$, and $35.1''$ and reach a 1-$\sigma$ depth of 1.6, 1.3 and 1.9 mJy at 250, 350 and 500 $\mu$m, respectively. We use the deblended SPIRE maps described in \citet{swinbank14}. These were obtained by constructing a catalog of IR-- and radio--bright galaxies based on ancillary multiwavelength data, in particular {\em Spitzer}/MIPS\,24$\mu$m, VLA, and the catalog of precisely located SMGs from ALESS. For each SPIRE band, a model of the image was created by assigning to each galaxy in the input catalog a SPIRE PSF scaled to a random flux between 0 and 1.3 times the brightest flux observed in the same region of the sky. This step was iterated until the model converged towards the observed map. For each sample of galaxies in our analysis ($K$-selected, sBzK, ERO, DRG), we remove all the contaminants by subtracting the best model of all the other (non colour selected) sources in the field. These `residual maps' are used as input images for our stacking routine. The stacking strategy follows that performed for the ALESS data. Here we focus on the $z>1$ subsamples. Our results are listed in Table~\ref{tab_spire} and shown in Fig.~\ref{fig_spire_ps}.

\section{Results}\label{sec_results}

\subsection{ALESS stacks}

Fig.~\ref{fig_stack} and Table \ref{tab_results} summarize the results of the stacking analysis applied to ALESS data. We also report the rms of the flux per pixel in the stacked maps (measured on the background at $>3''$ from the center), and the uncertainties estimated through the random stacks, as described in the previous section. These numbers allow us to gauge the robustness of a detection in our stacked images.

Clear detections are reported in most of the galaxy samples. The stacked flux of the $K_{\rm Vega}<20$ mag sample is likely dominated by a small subsample of bright sources at $z>1$: After removing the 10 sub-mm brightest sources, exceeding 1.8 mJy (i.e., 10\% of the sample), the stacked flux drops by a factor $\sim 3.4$. On the other hand, a positive, 5-$\sigma$ signal is reported also when we adopt median stacking instead of weighted averages. The stacked flux is $1.5$ times higher if we restrict our analysis only at the $z>1$ sources (we note that $\sim73$\% of the $K_{\rm Vega}<20$ mag galaxies at $z>1$ in the ALESS coverage are also classified as sBzK, EROs or DRGs). sBzK galaxies, EROs and DRGs all show clear detections ($>$10-$\sigma$) when the whole sample is considered. As we lower the 344 GHz flux cutoff to 1.2 mJy, DRGs still show significant detections ($\sim5$-$\sigma$), while detections at lower significance ($\sim3$-$\sigma$) are reported for sBzK and EROs. Median stacks performed without flux thresholds reveal clear detections in all the galaxy samples, suggesting that our results are robust against the contribution of bright outliers. The $z>1$ cutoff has marginal (if any) effect on these samples. When comparing the various samples of galaxies in our analysis, we find that, for any flux or redshift cut, sBzK and DRGs exhibit the brightest 344 GHz fluxes. This may be explained by the fact that the fraction of star-forming galaxies, as identified by the sBzK selection, is typically higher among DRGs ($\sim 35$\%) than among EROs \citep[$\sim 20$\%,][]{lane07} or $K$-selected galaxies \citep[$\sim 9$\%,][]{greve10}.

\subsection{{\em Herschel}/SPIRE stacks}\label{sec_spire}

Fig.~\ref{fig_spire_ps} and Table~\ref{tab_spire} summarize the results of the stacking analysis applied to SPIRE data. Clear detections are reported in all the bands and for all the galaxy samples. We use these measurements to constrain the SED of dust emission in color-selected galaxies at $z>1$. We fit the observed SEDs as modified black bodies $S_\nu \propto \nu^{\beta} B_\nu(T_{\rm dust})$, where $B_\nu(T)$ is the Planck function, $T_{\rm dust}=12-60$ K is the dust temperature, and $\beta=1.4-2.0$  sets the frequency dependence of the dust opacity \citep[e.g.,][]{beelen06,kelly12}. We find $\beta=1.6$ and $T_{\rm dust}\approx 30$ K in all the cases. These values are in agreement with the findings of similar studies: \citet{swinbank14} use combined MIPS 24$\mu$m, {\em Herschel}, ALESS, and VLA data in order to constrain the dust SED in ALESS SMGs, and find typical dust temperatures of $T_{\rm dust}=20-40$ K. \citet{bourne12} use {\em Herschel} Astrophysical Terahertz Large Area Survey \citep[H-ATLAS;][]{eales10} data at 250, 350 and 500 $\mu$m to constrain the dust SED of optically-selected star-forming galaxies at high-$z$. They find that blue and red galaxies are well described with a modified black body with $\beta\approx2$ and $T_{\rm dust}=10-30$ K (red galaxies having on average lower $T_{\rm dust}$ and $L_{\rm dust}$). \citet{elbaz11} use {\em Herschel} observations to infer typical SEDs of so-called main sequence and starbursting galaxies. They find an effective dust peak temperatures $T_{\rm eff}^{\rm peak}$ of 31 K for main sequence galaxies and 40 K for starbursts. 

IR luminosities are computed as the integral of the galaxy rest--frame spectral energy distribution between 8 and 1000 $\mu$m \citep[e.g.,][]{sanders03}. These luminosities are then converted into star formation rates by using: log SFR/(\Msun{}\,yr$^{-1}$) = log($1.3$) -- 10 + log ($L_{\rm IR}$/\Lsun{}) \citep[e.g.,][]{genzel10}. This conversion implicitly assumes a \citet{chabrier03} initial mass function. We find that SFRs in our samples range between 75 and 140 \Msun{}\,yr$^{-1}$ (see Table~\ref{tab_spire}). 

\begin{table*}
\begin{center}
\caption{Stacked SPIRE fluxes and estimates of dust temperature, IR luminosity and SFR.}\label{tab_spire}
\begin{tabular}{cccccccc}
\hline
 Sample  & $\langle z\rangle$ & $S_{\nu}$(250\,GHz) & $S_{\nu}$(350\,GHz) & $S_{\nu}$(500\,GHz) & $T_{\rm dust}$ & log\,$L_{\rm IR}$ & SFR \\
 $z>1$  &           & [mJy]  & [mJy]  & [mJy]   & [K] & [\Lsun{}] & [\Msun{}\,yr$^{-1}$] \\
 (1)     & (2)       &  (3)   & (4)    & (5)     & (6) &  (7)      &  (8)                 \\
\hline
$K_{\rm Vega}<20$  & 1.562 & $11.0\pm1.7$ & $12.0\pm2.1$ & $8.9\pm1.0$ & $32_{-2}^{+7}$ & $12.01_{-0.11}^{+0.15}$ & $130_{-30}^{+50}$ \\
sBzK	           & 1.896 & $ 7.7\pm1.1$ &  $7.6\pm1.3$ & $6.8\pm0.9$ & $29_{-4}^{+4}$ & $11.75_{-0.12}^{+0.09}$ &  $74_{-19}^{+17}$ \\
ERO	           & 1.502 & $12.2\pm1.7$ & $10.4\pm2.0$ & $7.6\pm1.0$ & $32_{-3}^{+8}$ & $12.03_{-0.14}^{+0.19}$ & $140_{-40}^{+70}$ \\
DRG	           & 1.792 & $10.8\pm1.6$ &  $9.6\pm1.7$ & $6.9\pm1.0$ & $30_{-4}^{+6}$ & $11.88_{-0.12}^{+0.15}$ & $100_{-20}^{+40}$ \\
\hline			
\vspace{-0.8mm}\\	
\end{tabular}		
\end{center}		
\end{table*} 		
\begin{figure}[t]
\includegraphics[width=0.99\columnwidth]{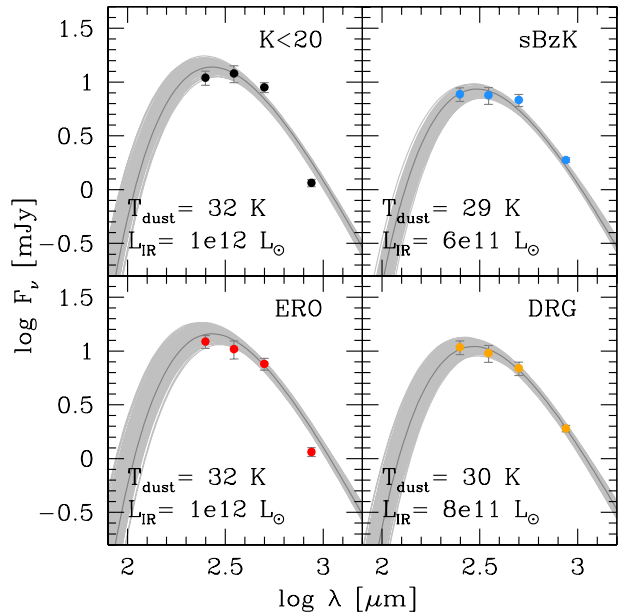}
\caption{Spectral Energy Distributions of $z>1$ color-selected galaxies in our analysis. Points (with error bars) show the {\em Herschel}/SPIRE and the ALESS fluxes obtained in our stacking analysis. The grey curves show the modified black body templates which best fit the data, and the range of models which are in good agreement (within 1-$\sigma$) with the observed constraints. Relevant best fit parameters are also reported.
}
\label{fig_spire}
\end{figure}

\subsection{Comparison with previous results at 870~$\mu$m}

Table~\ref{tab_results} (bottom) reports the stacked fluxes obtained in G10 for their sub-mm faint sample (roughly comparable with our $S_\nu$(344\,GHz)$<$3.6 mJy in terms of depth), and for their whole sample. We find a factor $\sim3$ brighter fluxes than G10 for all galaxy classes, if no $S_\nu$(344 GHz) cutoff is considered in our analysis. \citet{webb04} study the sub-mm properties of EROs using SCUBA. They find average 850 $\mu$m fluxes of $0.40\pm0.07$ mJy for galaxies selected by requiring ($I-K$)$_{\rm Vega}>4$, and $0.53\pm0.09$ mJy for galaxies with ($R-K$)$_{\rm Vega}>5.3$ (i.e., EROs in our classification). \citet{knudsen05} obtain a 850 $\mu$m average flux of $1.1\pm0.3$ mJy for DRGs, after stacking SCUBA observations of a cluster field (not corrected for lensing magnification). Similarly, by stacking SCUBA observations of 24$\mu$m-detected BzK galaxies,  \citet{daddi05} find an 850 $\mu$m average flux of $1.0\pm0.1$ mJy. Based on an independent stacking analysis on SCUBA observations of the Lockman hole and of the Subaru/{\em XMM-Newton} Deep Field, \citet{takagi07} report 850 $\mu$m fluxes of $0.52\pm0.19$ mJy, $0.52\pm0.16$ mJy and $0.3\pm0.3$ mJy for sBzK, EROs and DRGs.

\begin{figure}[t]
\includegraphics[width=0.99\columnwidth]{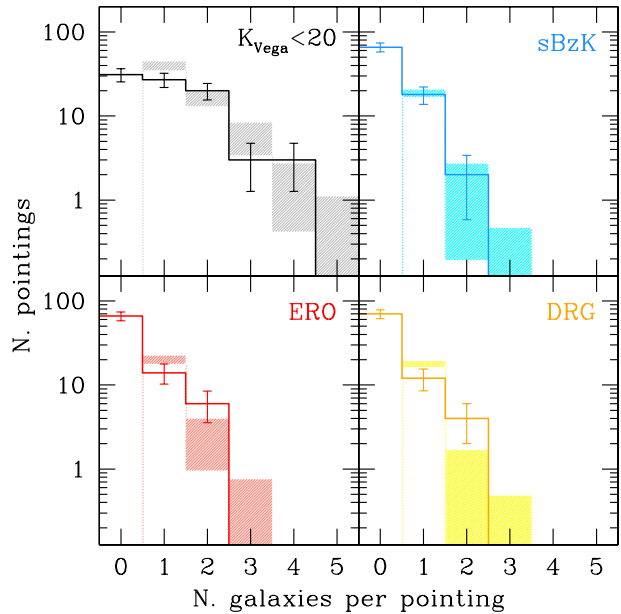}
\caption{Comparison between the observed number of galaxies per ALESS pointing (histograms) and the expectations from random sets of galaxies of the same class derived from the general field (shaded areas). Error bars show the poissonian uncertainties. The observed distributions are in agreement with those of the general sample, suggesting that ALESS sources have similar clustering properties as those in the whole MUSYC catalog.}
\label{fig_cluster}
\end{figure}
Overall, we conclude that we find brighter fluxes than most of the stacking studies performed so far at these wavelengths using single-dish data over larger fields. Possible explanations for this discrepancy are: 1) the discrepancy is an artifact, resulting by, e.g., the deblending procedures adopted in single-dish studies; 2) the discrepancy is real, i.e., the galaxies covered by ALESS pointings (defined to include the sub-mm brightest sources in the field) are intrinsically different from the general field covered in single-dish observations. The first scenario seems unlikely, given that various groups have run different analyses based on independent approaches and routines, and find (to first order) consistent results. In particular, \citet{daddi05} and \citet{takagi07} do not apply any deblending. 

In order to address the second scenario, we compare various properties of the galaxy samples considered in our analysis against the general field catalog and the LESS results from G10.

\begin{figure*}
\begin{center}
\includegraphics[width=0.89\textwidth]{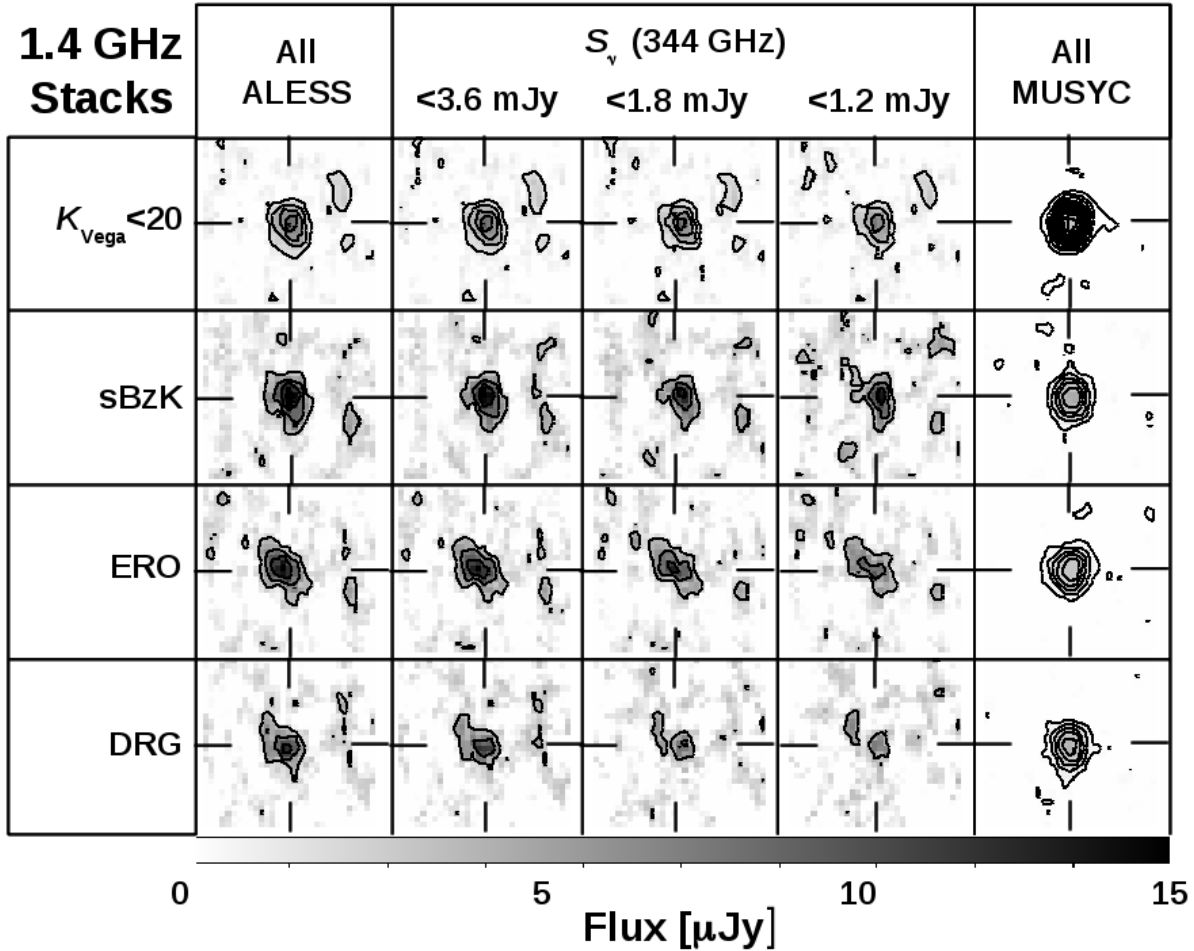}
\end{center}
\caption{Postage stamps of the 1.4 GHz median stacks of galaxies in ALESS, compared with the general MUSYC field. The first column refers to the stacks of the whole ALESS samples, while the second, third and fourth columns show the stack results of the samples after removing sources with $S_\nu$(344\,GHz)$>$3.6 mJy, 1.8 mJy and 1.2 mJy, respectively. The last column shows the stack results for the all the galaxies in the MUSYC sample. The gray scale is the same in all the panels, while contours are 2-$\sigma$ levels highlighting the significance of the stacked detections. Each postage stamp covers an area of $10''\times10''$. The radio flux decreases as we lower the cutoff in 344 GHz flux, as a result of the radio/IR correlation. The stacks from the general field show higher significance (thanks to the much larger sample sizes) but significantly lower fluxes than the sources encompassed by ALESS (see Table\,\ref{tab_radio}).}
\label{fig_radio_ps}
\end{figure*}

\begin{figure*}
\begin{center}
\includegraphics[width=0.89\textwidth]{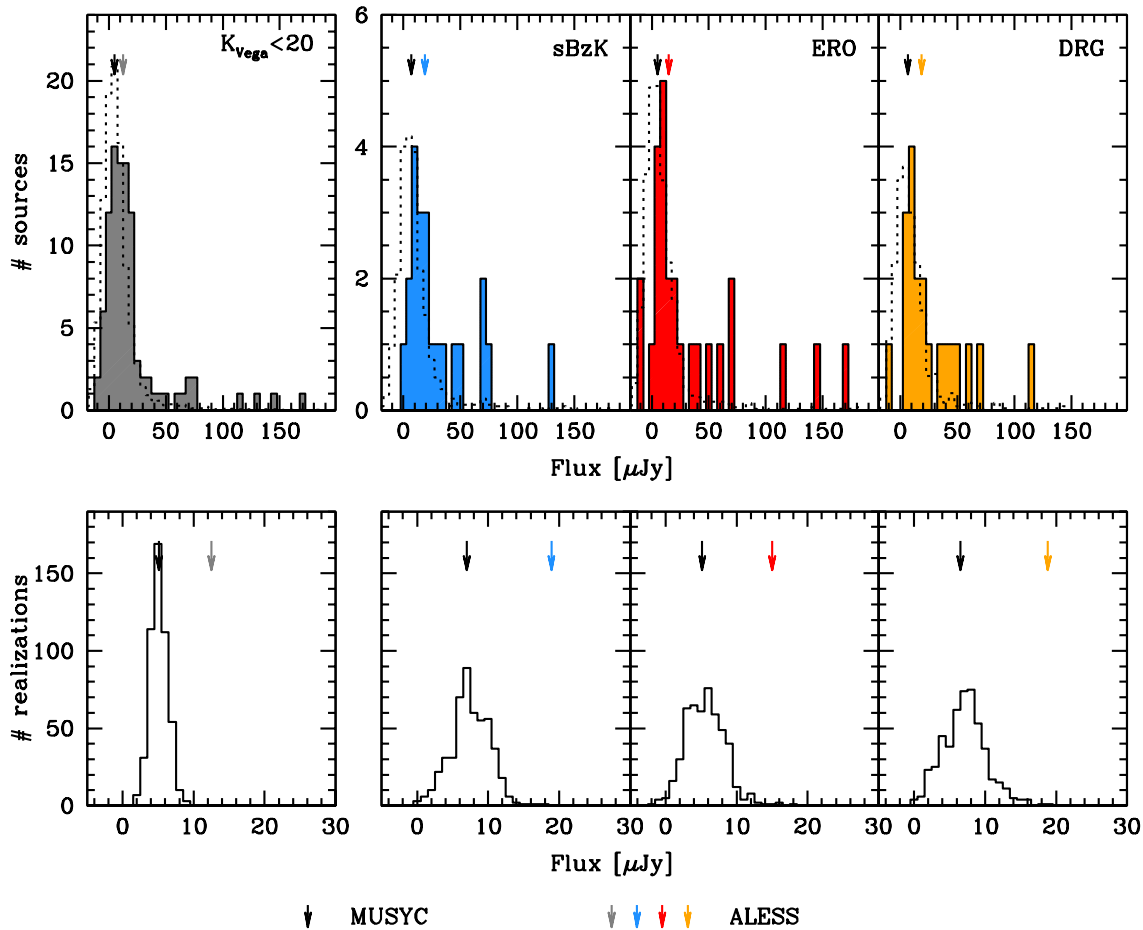}
\end{center}
\caption{{\em Top panels:} Comparison between the 1.4\,GHz fluxes of the sources in our analysis (shaded histograms) with respect to the general population in the MUSYC catalog (dotted histograms). Median fluxes are labeled with downward arrows. Galaxies covered in ALESS pointings are systematically brighter at 1.4\,GHz than the general population in the MUSYC field. {\em Bottom panels:} Distributions of the median 1.4\,GHz fluxes obtained by bootstrapping the MUSYC sample of galaxies, compared with the median values obtained for the ALESS galaxies (marked as arrows). The discrepancy is highly significant in all the galaxy color selections, suggesting that the discrepancy is real and does not result from an accidentaly radio-bright, randomly-selected subsample of typically fainter galaxies from the MUSYC catalog.}
\label{fig_radio}
\end{figure*}

\subsubsection{Redshift and NIR luminosity distributions}\label{sec_z_distr}

Fig.~\ref{fig_z_distr} compares the redshift and NIR luminosity distributions of the galaxies in our sample, with those of the general photometric catalog, scaled in order to have the same total number of sources.  Color-selected galaxies observed in ALESS tend to have slightly higher redshifts and brighter NIR luminosities. We use a Kolmogorov-Smirnov test to assess the significance of this discrepancy. The discrepancy is marginal for the redshift distributions of sBzK and pBzK galaxies, and is robust ($>3$-$\sigma$) for all the other distributions. In particular, the discrepancy in the redshift distributions is dominated by a small excess of ALESS--covered galaxies at $z\approx2$. Almost all the galaxies classified as sBzK, EROs, and DRGs in our analysis show $M_{\rm H}<-23$ mag, i.e., they are on average a factor $\sim2.5$ times brighter than their analogues in the general field. If $H$-band and 344 GHz luminosities are correlated in these sources \citep[similarly to what observed at radio wavelengths, see e.g.][]{dunne09}, then the difference observed in the $H$-band luminosity distributions would explain the discrepancy between our results and those of single-dish studies. 

\subsubsection{Clustering}\label{sec_cluster}

The ALESS pointings may be biased towards regions with most prominent overdensities of sub-mm bright galaxies. This is supported by various lines of evidence: 1) The brightest LESS sources appear split in multiple detections once observed at $\sim1.6''\times1.15''$ resolution \citep{karim13,hodge13}; 2) SMGs tend to lie in the progenitors of moderate to high mass groups of galaxies \citep[$M_{\rm halo} \sim 4\times10^{12}$ \Msun{}, see][]{hickox12}; 3) pairs or multiplets of EROs are often associated with sub-mm bright regions \citep{ivison02,webb04,chapman09,wardlow11}. \citet{aravena10} found significant overdensities of actively star-forming galaxies around three MAMBO-detected galaxies in the COSMOS field, although they do not find similar overdensities around other MAMBO sources in the same survey, suggesting that the occurrance of such structures around SMGs is about $\sim30$\%.

We test this scenario by comparing the clustering properties of galaxies in each class inside the ALESS coverage with those in the general photometric catalog. Fig.~\ref{fig_cluster} shows the number of ALESS pointings encompassing 0, 1, 2, \ldots{} color-selected galaxies. For each galaxy class we extract random, equally-sized sets of galaxies from the general field catalog, and we compute the number of galaxies of the same class that would be covered in ALESS--sized pointings centered on such a random sample. This process is repeated 50 times, allowing us to empirically evaluate the sample variance. Results of this Monte-Carlo test are shown in Fig.~\ref{fig_cluster} as shaded areas. In all the galaxy classes, we find consistency between the distributions of sources per pointing observed in the ALESS data and the ones of the simulations, suggesting that the sources covered in ALESS data show similar clustering properties compared to those of the general field. 

\subsubsection{Radio fluxes}\label{sec_radio}

\begin{table*}
\begin{center}
\caption{Summary of the 1.4\,GHz median stacking results.}\label{tab_radio}
\begin{tabular}{ccccccc}
\hline
Sample & All ALESS & \multicolumn{3}{c}{$S_\nu$(344\,GHz)} & All field  & ALESS excess \\
       &           & $<3.6$ mJy & $<1.8$ mJy & $<1.2$ mJy  &            &	       \\
       & [$\mu$Jy] & [$\mu$Jy]  & [$\mu$Jy]  & [$\mu$Jy]   & [$\mu$Jy]  &              \\
 (1)   & (2)       &  (3)       & (4)        & (5)         & (6)        & (7)	       \\
\hline
$K_{\rm Vega}<20$  & $10.0 \pm1.0 $ & $ 9.8 \pm1.0 $  & $ 9.1 \pm1.0 $ & $ 8.2 \pm1.0 $ & $ 3.94\pm0.16$  & $ 2.5\pm0.3 $ \\
sBzK	           & $17.9 \pm1.9 $ & $16.1 \pm2.0 $  & $14.5 \pm2.2 $ & $13.8 \pm2.2 $ & $ 5.3 \pm0.4 $  & $ 3.4\pm0.4 $ \\
ERO	           & $15.0 \pm1.8 $ & $14.8 \pm1.8 $  & $13.2 \pm1.8 $ & $10.0 \pm1.9 $ & $ 3.8 \pm0.3 $  & $ 3.9\pm0.6 $ \\
DRG	           & $13.0 \pm1.9 $ & $12.4 \pm2.1 $  & $ 9.1 \pm2.1 $ & $ 8.0 \pm2.3 $ & $ 4.8 \pm0.4 $  & $ 2.7\pm0.5 $ \\
\hline				 			    
\vspace{-0.8mm}\\
\end{tabular}
\end{center}
\end{table*} 

If we evaluate the contribution of the sources in our analysis to the Extragalactic Background Light (EBL) at 344 GHz, we obtain significantly higher surface brightnesses than expected. Our analysis covered $74 \times \pi (1.2 \times 17.3/2)^2$ arcsec$^2$ $\approx$ 0.00193 deg$^2$. If we divide the total flux from $K_{\rm Vega}<20$ mag galaxies, sBzK, EROs and DRGs by this area, we obtain: 40.4 Jy deg$^{-2}$, 21.4 Jy deg$^{-2}$, 15.0 Jy deg$^{-2}$, and 18.3 Jy deg$^{-2}$ respectively, i.e., $5-10$ times higher than what was reported by G10 for the whole LESS field, and close to the total EBL light ($44\pm15$ Jy deg$^{-2}$, see \S5.4 in G10; note however that there is substantial overlap between our galaxy samples as defined by the adopted color cuts). Such a discrepancy suggests that, by survey design, ALESS pointings encompass regions of the sky which are intrinsically brighter at sub-mm wavelengths than the general field. 

In order to test this scenario, we compare the radio (1.4\,GHz) fluxes of the sources covered by ALESS pointings against the sources in the general field. Our test relies on the observed 1.4\,GHz -- IR luminosity relation \citep{condon82,helou85,yun01,garrett02,ivison10,sargent10}. Given the small differences in the redshift distributions of the photometric sample within and outside the ALESS coverage (as discussed in \S\ref{sec_z_distr}), if galaxies covered in ALESS pointings are intrinsically brighter at 344\,GHz than the galaxies in the general MUSYC field, we expect a similar difference to be observed also at 1.4\,GHz. We base our comparison on the 1.4\,GHz map of the ECDFS obtained by \citet{miller08} \citep[see also][]{miller13}. The map was obtained with the Very Large Array in extended (A) configuration, yielding a spatial resolution of $2.8''\times1.6''$ over a $32'\times32'$ wide region. The typical rms of the mosaic is $\sim7$ $\mu$Jy\,beam$^{-1}$. 

In Fig.~\ref{fig_radio_ps} we show the stacks of the 1.4\,GHz images of the sources in our sample. We perform radio stacks also for the sub-samples obtained with various sub-mm flux thresholds. The results are compared with the median stacks obtained for the general population of color-selected galaxies in the field. We perform both median and average stacks. The latter however are dominated by the contribution of a few, very bright outliers (most likely, radio-loud active galactic nuclei). Because of this, we focus only on median stacks. All the galaxy samples ($K_{\rm Vega}<20$ mag, sBzK, EROs, and DRGs) show clear 1.4\,GHz detections in the stacks. The significance of such detections drops as we remove sub-mm bright sources, as a result of the IR--radio luminosity relation. The detection is only marginal (3.5-$\sigma$) for DRGs as we apply the most aggressive sub-mm flux cut ($S_\nu$(344\,GHz) $<1.2$ mJy). Remarkably, the stacks over the entire sample show considerably fainter 1.4\,GHz fluxes, by a factor $\sim3$ compared with the uncut ALESS sample (see Table~\ref{tab_radio}). The latter stacked fluxes are slightly lower but still in agreement with the results of similar stacking studies at 1.4 GHz by \citet{dunne09}, which are clearly inconsistent with the values obtained for the ALESS--covered samples. The discrepancy in the stacked fluxes within and outside ALESS pointings is still significant even if we consider only the sources which are not individually detected in the ALESS observations. This suggests that the diffence in sub-mm and radio fluxes between ALESS-covered galaxies and the general sample extends to a number of galaxies in the close neighborhood of sub-mm galaxies.

In order to further assess the robustness of this result, in Fig.~\ref{fig_radio} we show the distribution of the 1.4\,GHz fluxes of field sources within and outside the ALESS pointings. Sources in the ALESS coverage are systematically brighter. We compute median fluxes of random sub-samples of the field galaxies, requiring the same sample size as the one covered in ALESS pointings. This bootstrapping approach allows us to evaluate whether the discrepancy between the general field and ALESS-covered galaxies can be explained as an accidentally-bright realization drawn from a typically fainter population. This test (shown in the bottom panels of Fig.~\ref{fig_radio}) shows that the observed median values of the 1.4\,GHz stacks of ALESS-covered galaxies represent $>4$-$\sigma$ outliers compared with the distribution of median stack values derived from bootstrapping the field sample. This result is confirmed via a Kolmogorov-Smirnov test, which rule out that the distributions of 1.4\,GHz fluxes of the ALESS-covered galaxies and of the general field population are drawn from the same parent distribution at $>4$-$\sigma$ for all the color selections.

\subsubsection{Physical association vs.~chance projection}

Our analysis so far has shown the following: 1) we observe a discrepancy in the stacked fluxes of galaxies within ALESS pointings and the results from similar studies based on single-dish observations over larger fields. 2) The discrepancy is likely driven by different sub-mm properties of galaxies (e.g., galaxies around SMGs tend to be brighter at sub-mm wavelengths than typical field galaxies with similar NIR luminosities). 3) This hypothesis is confirmed by a similar discrepancy observed at 1.4\,GHz, even after removing galaxies individually detected at sub-mm wavelengths. We therefore conclude that galaxies in the vicinity of SMGs (on sky) tend to be brighter than those in the general field. 

We now investigate whether this effect is due to chance alignments (i.e., ALESS pointings cover regions of the sky that show projected overdensities of sub--mm bright galaxies) or to physical associations (i.e., the sub-mm bright sources belong to physical overdensities around individually--detected ALESS SMGs). From each of our galaxy samples, we consider as physically associated all the galaxies (including the SMGs) with a photometric redshift \citep[from][]{simpson14} consistent with the photometric redshift of the nearby ($<200$\,kpc in terms of projected distance) SMG. 
These galaxies are highlighted with thick lines in Fig.~\ref{fig_histo}. Galaxies physically associated with an SMG seem to show brighter 870\,$\mu$m fluxes than the general color-selected samples covered by ALESS observations, although this is only tentative due to the limited sample sizes. If confirmed, this result would suggest that the discrepancy between the sub-mm (and radio) fluxes of color-selected galaxies inside ALESS pointings, and those in the general field, is intrinsic in the properties of galaxies, and not simply due to chance superposition. In this scenario, galaxies spatially close to SMGs would show brighter sub-mm fluxes than similar galaxies in the field. We note however that the uncertainties in the photometric redshifts used in this test are large. This implies that the ``physical association'' flag used here may be a by-product of the similar redshift distributions of SMGs and our color-selected galaxies, rather than the result of real physical connection.

\subsection{Star formation rate density}

\begin{figure}
\includegraphics[width=0.99\columnwidth]{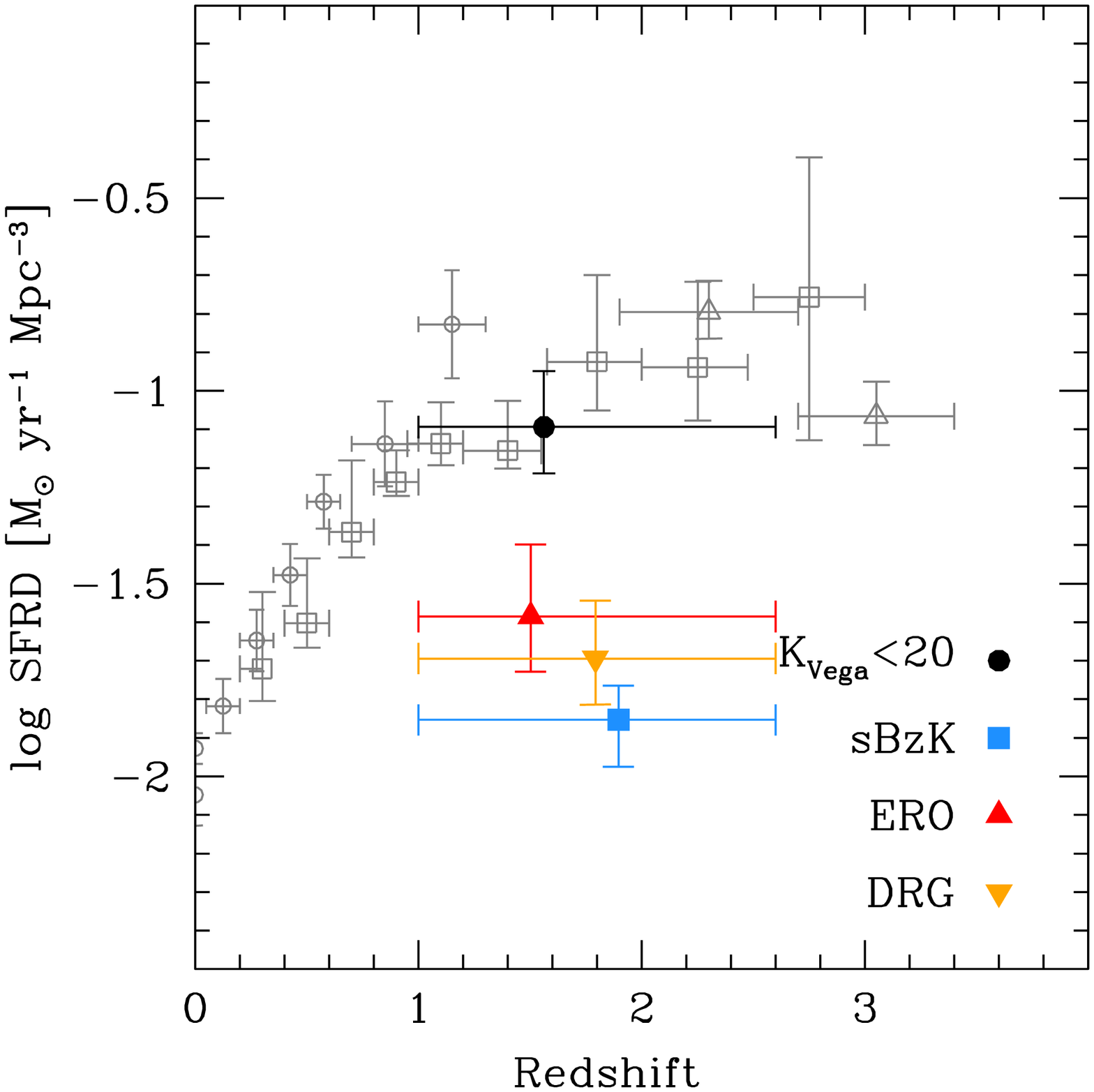}
\caption{Cosmic star formation rate density as a function of redshift. Grey open points are IR/radio-based SFR density estimates from the literature: squares from \citet{karim11}, circles from \citet{rujopakarn10} and triangles from \citet{reddy08}, homogenized to a \citet{chabrier03} initial mass function. Filled symbols show the results from our stacking analysis based on galaxies observed in ALESS pointings. Error bars account for the formal uncertainties in the stacked fluxes, but not for the uncertainties due to model assumptions. In order to infer SFR densities, we have assumed a modified black body template scaled in order to match the 344 GHz fluxes measured in our study for $z>1$ galaxies. The SFRD of $K$-selected galaxies is in agreement with previous works in the literature, although we remark that we did not correct for the excess in IR emission observed in sources within ALESS pointings. sBzK, EROs, and DRGs account for $\sim17$\%, $\sim33$\%, and $\sim25$\% of the total SFRD from $K$-selected galaxies, respectively.
}
\label{fig_madau}
\end{figure}

In Fig.~\ref{fig_madau} we use the estimated SFRs for the $z>1$ galaxies computed in \S\ref{sec_spire} in order to put constraints on the cosmic SFR density (SFRD) as a function of redshift. It is claimed that SFRD smoothly increased from very high-redshift until $z=1-2$, when it reached its peak \citep[$\sim 0.2$ \Msun{}\,yr$^{-1}$\,Mpc$^{-3}$, assuming a][initial mass function]{chabrier03}. This is usually referred to as the `epoch of galaxy assembly', when roughly half of the stars in the universe were formed. At more recent cosmic times, the SFRD declined by more than an order of magnitude between $z\sim 1$ and $z$=0 \citep{lilly95,madau96,hopkins06,reddy08,rujopakarn10,bouwens10,karim11}. We compute SFRDs associated to our sample of galaxies by multiplying the average SFRs by the number of galaxies in each sample, and dividing the resulting `total' SFRs by the volume sampled in our analysis. As described in \S\ref{sec_musyc}, the photometric catalog is highly complete at $K_{\rm Vega}<20$ mag. At this flux threshold, we sample galaxies up to $z$=2.6 (see Fig.~\ref{fig_z_distr}). We thus restrict our analysis to the $1<z<2.6$ range, where color selection is most effective. Our analysis covered a total of $7.0$ square arcmin. The corresponding comoving volume is 34000 Mpc$^3$. Finally, we scale our SFRD estimates down in order to account for the excess in ALESS fluxes with respect to the general field (see Table \ref{tab_radio}).

We compare our results with those of other studies available in the literature, based on IR/radio SFRD measurements \citep{reddy08,rujopakarn10,karim11}. The SFRD derived in \citet{rujopakarn10} are scaled down by a factor 1.65 in order to account for the different initial mass function assumption. 

Our estimates of the SFRD for $K$-selected galaxies are in broad agreement with previous work in the literature ($\approx 0.08$ \Msun{}\,yr$^{-1}$\,Mpc$^{-3}$). The other subsamples of galaxies (sBzK, EROs, and DRGs) account for $\sim17$\%, $\sim33$\%, and $\sim25$\% of the total SFRD from $K$-selected galaxies, respectively (note that these samples have substantial overlap, see \S\ref{sec_musyc}).

\section{Conclusions}\label{sec_conclusions}

We present a stacking analysis of ALMA interferometric 344 GHz continuum observations of high-$z$ galaxies as part of the ALESS survey. We base our stacking approach on the photometric optical/NIR catalog of field galaxies in the ECDFS. Based on their fluxes and colors, we select four classes of galaxies: $K_{\rm Vega}<20$ mag, sBzK, EROs and DRGs. We find that:
\begin{itemize}
\item[{\em i-}] Clear ($>$10-$\sigma$) detections are reported for all the galaxy classes, independently of the averaging algorithm (weighted averages vs median stacks). The detections are robust also after excluding sources bright enough to be individually detected in the original single-dish LESS observations.
\item[{\em ii-}] The detection of 344 GHz flux in $K$-selected galaxies is dominated by their high-$z$ sub-sample. Half of the $K_{\rm Vega}<20$ mag galaxies lie at $z<1$ and show significantly fainter 344 GHz emission than the subsample at $z>1$. In particular, a few, very bright galaxies dominate the flux in the stacked maps (although a clear detection is reported also when median stacks are considered).
\item[{\em iii-}] Color-selected galaxies (sBzK, EROs, DRGs) are detected in our stacking analysis at 344 GHz, even after excluding sources which would be individually detected in the original LESS observations or in the ALESS maps.
\item[{\em iv-}] We perform a similar stacking analysis on {\em Herschel}/SPIRE maps of the ECDFS, and combine these findings with our ALESS results, in order to constrain the shape of the dust SED in color-selected galaxies at $z>1$. We find that the IR emission of these sources is well described by a modified black body with $\beta=1.6$ and $T_{\rm dust}\approx30$ K. We infer IR luminosities of $L_{\rm IR}=(5-11) \times 10^{11}$ \Lsun{}, and associated SFRs of 75--140 \Msun{} yr$^{-1}$.
\item[{\em v-}] We find brighter 870\,$\mu$m fluxes than previously reported in similar stacking studies of this (G10) or other regions of the sky, based on single-dish observations at the same frequency. A similar discrepancy is observed if we apply our stacking approach to 1.4\,GHz observations of the ECDFS. 
\item[{\em vi-}] If we limit our analysis to sources with photometric redshifts consistent with a nearby SMG, we find tentative evidence that these galaxies are intrinsically brighter at 870\,$\mu$m than the remainder population. If confirmed, this result may be interpreted as a significant contribution of pairs of interacting galaxies to the SMG population, as has been suggested by theoretical models \citep[e.g.][]{baugh05}. However, significantly higher precision in the redshift estimates, and larger samples, are mandatory in order to support this scenario.
\item[{\em vii-}] When we place these SFR estimates into a cosmological context, we find that color--selected galaxies contribute to one third -- one sixth of the cosmic star formation rate density at $1<z<2.6$.
\end{itemize}
Our study demonstrates the power of stacking analyses applied on interferometric data at sub-mm wavelengths in unveiling the properties of star-forming galaxies at high-$z$. Our upcoming Cycle 1 ALMA observations of the same fields will allow us to individually detect the sources that can currently only be detected through stacking analysis, and will allow us to further push down our stacking sensitivity limits.
 
\section*{Acknowledgments}
We thank the anonymous referee for her/his useful comments. This paper makes use of the following ALMA data: ADS/JAO.ALMA\#2011.1.00294.S. ALMA is a partnership of ESO (representing its member states), NSF (USA) and NINS (Japan), together with NRC (Canada) and NSC and ASIAA (Taiwan), in cooperation with the Republic of Chile. The Joint ALMA Observatory is operated by ESO, AUI/NRAO and NAOJ. This publication also makes use of data acquired with Europear Southern Observatories VLT under program ID 183.A-0666. This publication is based on data acquired with the APEX under programme IDs 078.F-9028(A), 079.F-9500(A), 080.A-3023(A) and 081.F-9500(A). APEX is a collaboration between the Max-Planck-Institut f\"{u}r Radioastronomie, the European Southern Observatory and the Onsala Space Observatory. Support for RD was provided by the DFG priority program 1573 ``The physics of the interstellar medium''. IRS acknowledges support from STFC (ST/IOO1573/1), a Leverhulme Fellowship, the ERC Advanced Investigator programme DUSTYGAL and a Royal Society/Wolfson Merit Award. AMS gratefully acknowledges an STFC Advanced Fellowship (ST/H005234/1). RI acknowledges the ERC AdG programme COSMICISM. AK acknowledges support from STFC as well as from the Collaborative Research Council 956 funded by the Deutsch Forschungsgemeinschaft (DFG). KK thanks the Swedish Research Council for support (grant 621-2011-5372). The data employed in this analysis are available from the ALMA and ESO archives.

\label{lastpage}


\begin{thebibliography}{99}
\bibitem[\protect\citeauthoryear{Aleksi\'{c} et al.}{2011}]{aleksic11} Aleksi\'{c} J., Antonelli L.A., Antoranz P., Backes M., Baixeras C., Barrio J.A., Bastieri D., Becerra Gonz\'{a}lez J., et al. 2011, ApJ, 729, 115
\bibitem[\protect\citeauthoryear{Aravena et al.}{2010}]{aravena10} Aravena M., Bertoldi F., Carilli C., Schinnerer E., McCracken H.J., Salvato M., Riechers D., Sheth K., et al. 2010, ApJ, 708, L36
\bibitem[\protect\citeauthoryear{Austermann et al.}{2010}]{austermann10} Austermann J.E., Dunlop J.S., Perera T.A., Scott K.S., Wilson G.W., Aretxaga I., Hughes D.H., Almaini O., et al. 2010, MNRAS, 401, 160
\bibitem[\protect\citeauthoryear{Baugh et al.}{2005}]{baugh05} Baugh C.M., Lacey C.G., Frenk C.S., Granato G.L., Silva L., Bressan A.J., Cole S., 2005, MNRAS, 356, 1191
\bibitem[\protect\citeauthoryear{Barger et al.}{1999}]{barger99} B
\bibitem[\protect\citeauthoryear{Beelen et al.}{2006}]{beelen06} Beelen A., Cox P., Benford D.J., Dowell C.D., Kov\'{a}cs A., Bertoldi F., Omont A., Carilli C.L., 2006, ApJ, 642, 694
\bibitem[\protect\citeauthoryear{Bell}{2003}]{bell03} Bell E.F., 2003, ApJ, 586, 794
\bibitem[\protect\citeauthoryear{Blain et al.}{2002}]{blain02} Blain A.W., Smail I., Ivison R.J., Kneib J.-P., Frayer D.T., 2002, Phys. Rev., 369, 111
\bibitem[\protect\citeauthoryear{Borys et al.}{2003}]{borys03}  Borys C., Chapman S., Halpern M., Scott D., 2003, MNRAS, 344, 385
\bibitem[\protect\citeauthoryear{Bourne et al.}{2012}]{bourne12} Bourne N., Maddox S.J., Dunne L., Auld R., Baes M., Baldry I.K., Bonfield D.G., Cooray A., et al. 2012, MNRAS, 421, 3027
\bibitem[\protect\citeauthoryear{Bouwens et al.}{2010}]{bouwens10} Bouwens R.J., Illingworth G.D., Oesch P.A., Stiavelli M., van Dokkum P., Trenti M., Magee D., Labb\'{e} I., Franx M., et al. 2010, ApJ, 709, L133
\bibitem[\protect\citeauthoryear{Boyle et al.}{2007}]{boyle07} Boyle B.J., Cornwell T.J., Middelberg E., Norris R.P., Appleton P.N., Smail I., 2007, MNRAS, 376, 1182
\bibitem[\protect\citeauthoryear{Brinchmann et al.}{2004}]{brinchmann04} Brinchmann J., Charlot S., White S.D.M., Tremonti C., Kauffmann G., Heckman T., Brinkmann J., 2004, MNRAS, 351, 1151
\bibitem[\protect\citeauthoryear{Chabrier}{2003}]{chabrier03} Chabrier G., 2003, PASP, 115, 763
\bibitem[\protect\citeauthoryear{Chapman et al.}{2003}]{chapman03} Chapman S.C., Blain A.W., Ivison R.J., Smail I.R., 2003, Nature, 422, 695
\bibitem[\protect\citeauthoryear{Chapman et al.}{2005}]{chapman05} Chapman S.C., Blain A.W., Smail I., Ivison R.J., 2005, ApJ, 622, 772
\bibitem[\protect\citeauthoryear{Chapman et al.}{2009}]{chapman09} Chapman S.C., Blain A., Ibata R., Ivison R.J., Smail I., Morrison G., 2009, ApJ, 691, 560
\bibitem[\protect\citeauthoryear{Chaudhary et al.}{2012}]{chaudhary12} Chaudhary P., Brusa M., Hasinger G., Merloni A., Comastri A., Nandra K., 2012, A\&A, 537, 6
\bibitem[\protect\citeauthoryear{Condon et al.}{1982}]{condon82} Condon J.J., Condon M.A., Gisler G., Puschell J.J., 1982, ApJ, 252, 102
\bibitem[\protect\citeauthoryear{Condon}{1992}]{condon92} Condon J.J., 1992, ARA\&A, 30, 575
\bibitem[\protect\citeauthoryear{Coppin et al.}{2006}]{coppin06} Coppin K., Chapin E.L., Mortier A.M.J., Scott S.E., Borys C., Dunlop J.S., Halpern M., Hughes D.H., et al. 2006, MNRAS, 372, 1621
\bibitem[\protect\citeauthoryear{Da Cunha et al.}{2012}]{dacunha12} Da Cunha E., Walter F., Decarli R., Bertoldi F., Carilli C., Daddi E., Elbaz D., Ivison R., Maiolino R., et al. 2012, submitted
\bibitem[\protect\citeauthoryear{Daddi et al.}{2004}]{daddi04} Daddi E., Cimatti A., Renzini A., Fontana A., Mignoli M., Pozzetti L., Tozzi P., Zamorani G., 2004, ApJ, 618, 23
\bibitem[\protect\citeauthoryear{Daddi et al.}{2005}]{daddi05} Daddi E., Dickinson M., Chary R., Pope A., Morrison G., Alexander D.M., Bauer F.E., Brandt W.N., et al. 2005, ApJ, 626, 680
\bibitem[\protect\citeauthoryear{Daddi et al.}{2007}]{daddi07} Daddi E., Dickinson M., Morrison G., Chary R., Cimatti A., Elbaz D., Frayer D., Renzini A., et al. 2007, ApJ, 670, 156
\bibitem[\protect\citeauthoryear{Damen et al.}{2011}]{damen11} Damen M., Labb\'{e} I., van Dokkum P.G., Franx M., Taylor E.N., Brandt W.N., Dickinson M., Gawiser E., Illingworth G.D., Kriek M., Marchesini D., Muzzin A., Papovich C., Rix H.-W., 2011, ApJ, 727, 1
\bibitem[\protect\citeauthoryear{Dole et al.}{2006}]{dole06} Dole H., Lagache G., Puget J.-L., Caputi K.I., Fern\'{a}ndez-Conde N., Le Floc'h E., Papovich C., P\'{e}rez-Gonz\'{a}lez P.G., et al. 2006, A\&A, 451, 417
\bibitem[\protect\citeauthoryear{Dunne et al.}{2009}]{dunne09} Dunne L., Ivison R.J., Maddox S., Cirasuolo M., Mortier A.M., Foucaud S., Ibar E., Almaini O., Simpson C., McLure R., 2009, MNRAS, 394, 3
\bibitem[\protect\citeauthoryear{Eales et al.}{2010}]{eales10} Eeles S., et al. 2010, A\&A, 518, L23
\bibitem[\protect\citeauthoryear{Elbaz et al.}{2011}]{elbaz11} Elbaz D., Dickinson M., Hwang H.S., D\'{i}az-Santos T., Magdis G., Magnelli B., Le Borgne D., Galliano F., et al. 2011, A\&A, 533, 119
\bibitem[\protect\citeauthoryear{Elston et al.}{1988}]{elston88} Elston R., Rieke G.H., Rieke M.J., 1988, ApJ, 331, L77
\bibitem[\protect\citeauthoryear{Fabello et al.}{2011}]{fabello11} Fabello S., Catinella B., Giovanelli R., Kauffmann G., Haynes M.P., Heckman T.M., Schiminovich D., 2011, MNRAS, 411, 993
\bibitem[\protect\citeauthoryear{Franx et al.}{2003}]{franx03} Franx M., Labb\'{e} I., Rudnick G., van Dokkum P.G., Daddi E., F\"{o}rster Schreiber N.M., Moorwood A., Rix H.-W., et al. 2003, ApJ, 587, L79
\bibitem[\protect\citeauthoryear{Garrett}{2002}]{garrett02} Garrett M.A., 2002, A\&A, 384, L19
\bibitem[\protect\citeauthoryear{Gawiser et al.}{2006}]{gawiser06} Gawiser E., van Dokkum P.G., Herrera D., Maza J., Castander F.J., Infante L., Lira P., Quadri R., et al. 2006, ApJ Suppl., 162, 1
\bibitem[\protect\citeauthoryear{Genzel et al.}{2010}]{genzel10} Genzel R., et al. 2010, MNRAS, 407, 2091
\bibitem[\protect\citeauthoryear{George et al.}{2012}]{george12} George M.R., Leauthaud A., Bundy K., Finoguenov A., Ma C.-P., Rykoff E.S., Tinker J.L., Wechsler R.H., et al. 2012, ApJ, 757, 2
\bibitem[\protect\citeauthoryear{Giacconi et al.}{2001}]{giacconi01} Giacconi R., Rosati P., Tozzi P., Nonino M., Hasinger G., Norman C., Bergeron J., Borgani S., Gilli R., et al. 2001, ApJ, 551, 624
\bibitem[\protect\citeauthoryear{Gonz\'{a}lez et al.}{2012}]{gonzales12} Gonz\'{a}lez V., Bouwens R.J., Labb\'{e} I., Illingworth G., Oesch P., Franx M., Magee D., 2012, ApJ, 755, 148
\bibitem[\protect\citeauthoryear{Greve et al.}{2010}]{greve10} Greve T., et al. 2010, ApJ, 719, 483
\bibitem[\protect\citeauthoryear{Hatsukade et al.}{2010}]{hatsukade10} Hatsukade B., Iono D., Akiyama T., Yoshikawa, M., Dunlop J.S., Ivison R.J., Peck A.B., Ikarashi S., et al. 2010, ApJ, 711, 974
\bibitem[\protect\citeauthoryear{Helou et al.}{1985}]{helou85} Helou G., Soifer B.T., Rowan-Robinson M., 1985, ApJ, 298, L7
\bibitem[\protect\citeauthoryear{Hickox et al.}{2012}]{hickox12} Hickox R.C., Wardlow J.L., Smail I., Myers A.D., Alexander D.M., Swinbank A.M., Danielson A.L.R., Stott J.P., et al. 2012, MNRAS, 421, 284
\bibitem[\protect\citeauthoryear{Hodge et al.}{2008}]{hodge08} Hodge J.A., Becker R.H., White R.L., de Vries W.H., 2008, AJ, 136, 1097
\bibitem[\protect\citeauthoryear{Hodge et al.}{2009}]{hodge09} Hodge J.A., Zeimann G.R., Becker R.H., White R.L., 2009, AJ, 138, 900
\bibitem[\protect\citeauthoryear{Hodge et al.}{2013}]{hodge13} Hodge J.A., Karim A., Smail I., Swinbank A.M., Walter F., Biggs A.D., Ivison R.J., Wei\ss{} A., et al. 2013, ApJ, 768, 91
\bibitem[\protect\citeauthoryear{Hopkins \& Beacom}{2006}]{hopkins06} Hopkins A.M. \& Beacom J.F., 2006, ApJ, 651, 142
\bibitem[\protect\citeauthoryear{Hsieh et al.}{2012}]{hsieh12} Hsieh B.-C., Wang W.-H., Hsieh C.-C., Lin L., Yan H., Lim J.,
Ho P. T. P., 2012, ApJS, 203, 23
\bibitem[\protect\citeauthoryear{Ivison et al.}{2002}]{ivison02} Ivison R.J., Greve T.R., Smail I., Dunlop J.S., Roche N.D., Scott S.E., Page M.J., Stevens J.A., Almaini O., Blain A.W., et al. 2002, MNRAS, 337, 1
\bibitem[\protect\citeauthoryear{Ivison et al.}{2007}]{ivison07} Ivison R.J., Greve T.R., Dunlop J.S., Peacock J.A., Egami E., Smail I., Ibar E., van Kampen E., Aretxaga I., Babbedge T., et al. 2007,  MNRAS, 380, 199
\bibitem[\protect\citeauthoryear{Ivison et al.}{2010}]{ivison10} Ivison R.J., Magnelli B., Ibar E., Andreani P., Elbaz D., Altieri B., Amblard A., Arumugam V., Auld R., Aussel H., et al. 2010, A\&A, 518, L31
\bibitem[\protect\citeauthoryear{Karim et al.}{2011}]{karim11} Karim A., Schinnerer E., Mart\'{\i}nez-Sansigre A., Sargent M.T., van der Wel A., Rix H.-W., Ilbert O., Smol\v{c}i\'{c} V., et al. 2011, ApJ, 730, 61
\bibitem[\protect\citeauthoryear{Karim et al.}{2013}]{karim13} Karim A., Swinbank A.M., Hodge J.A., Smail I., Walter F., Biggs A.D., Simpson J.M., Danielson A.L.R., et al. 2013, MNRAS, 432, 2
\bibitem[\protect\citeauthoryear{Kelly et al.}{2012}]{kelly12} Kelly B.C., Shetty R., Stutz A.M., Kauffmann J., Goodman A.A., Launhardt R., 2012, ApJ, 752, 55
\bibitem[\protect\citeauthoryear{Kennicutt}{1998}]{kennicutt98} Kennicutt R.C., 1998, ARA\&A, 36, 189
\bibitem[\protect\citeauthoryear{Kewley et al.}{2001}]{kewley01} Kewley L.J., Dopita M.A., Sutherland R.S., Heisler C.A., Trevena J., 2001, ApJ, 556, 121
\bibitem[\protect\citeauthoryear{Knudsen et al.}{2005}]{knudsen05} Knudsen K.K., van der Werf P., Franx M., F\"{o}rster Schreiber N.M., van Dokkum P.G., Illingworth G.D., Labb\'{e} I., Moorwood A., et al. 2005, ApJ, 632, L9
\bibitem[\protect\citeauthoryear{Knudsen et al.}{2009}]{knudsen09} Knudsen K.K., Neri R., Kneib J.-P., van der Werf P.P., 2009, A\&A, 496, 45
\bibitem[\protect\citeauthoryear{Kurczynzki \& Gawiser}{2010}]{kurczynski10} Kurczynski P., Gawiser E., 2010, AJ, 139, 1592
\bibitem[\protect\citeauthoryear{Lane et al.}{2007}]{lane07} Lane K.P., Almaini O., Foucaud S., Simpson C., Smail I., McLure R.J., Conselice C.J., Cirasuolo M., et al. 2007, MNRAS, 379, L25
\bibitem[\protect\citeauthoryear{Leroy et al.}{2012}]{leroy12} Leroy A.K., Bigiel F., de Blok W.J.G., Boissier S., Bolatto, A., Brinks E., Madore B., Munoz-Mateos J.-C., Murphy E., et al. 2012, AJ, 144, 3
\bibitem[\protect\citeauthoryear{Lilly et al.}{1995}]{lilly95} Lilly S.J., Tresse L., Hammer F., Crampton D., Le F\`{e}vre O., 1995, ApJ, 455, 108
\bibitem[\protect\citeauthoryear{Madau et al.}{1996}]{madau96} Madau P., Ferguson H.C., Dickinson M.E., Giavalisco M., Steidel C.C., Fruchter A., 1996, MNRAS, 283, 1388
\bibitem[\protect\citeauthoryear{Matsuda et al.}{2012}]{matsuda12} Matsuda Y., Yamada T., Hayashino T., Yamauchi R., Nakamura Y., Morimoto N., Ouchi M., Ono Y., Umemura M., Mori M., 2012, MNRAS, 425, 878
\bibitem[\protect\citeauthoryear{Miller et al.}{2008}]{miller08} Miller N.A., Fomalont E.B., Kellermann K.I., Mainieri V., Norman C., Padovani P., Rosati P., Tozzi P., 2008, ApJS, 179, 114
\bibitem[\protect\citeauthoryear{Miller et al.}{2013}]{miller13} Miller N.A.,   Bonzini M., Fomalont E.B., Kellermann K.I., Mainieri V., Padovani P., Rosati P., Tozzi P., Vattakunnel S., 2013, ApJS, 205, 13
\bibitem[\protect\citeauthoryear{Moncelsi et al.}{2011}]{moncelsi11} Moncelsi L., Ade P.A.R., Chapin E.L., Cortese L., Devlin M.J., Dye S., Eales S., Griffin M., et al. 2011, ApJ, 727, 83
\bibitem[\protect\citeauthoryear{Murphy et al.}{2012}]{murphy12} Murphy E.J., Bremseth J., Mason B.S., Condon J.J., Schinnerer E., Aniano G., Armus L., Helou G., et al. 2012, ApJ, 761, 97
\bibitem[\protect\citeauthoryear{Oke}{1974}]{oke74} Oke J.B., 1974, ApJ Suppl., 27, 21
\bibitem[\protect\citeauthoryear{Oliver et al.}{2012}]{oliver12} Oliver S.J., Bock J., Altieri B., Amblard A., Arumugam V., Aussel H., Babbedge T., Beelen A., et al. 2012, MNRAS, 424, 1614
\bibitem[\protect\citeauthoryear{Reddy et al.}{2008}]{reddy08} Reddy N.A., Steidel C.C., Pettini M., Adelberger K.L., Shapley A.E., Erb D.K., Dickinson M., 2008, ApJS, 175, 48
\bibitem[\protect\citeauthoryear{Rujopakarn et al.}{2010}]{rujopakarn10} Rujopakarn W., Eisenstein D.J., Rieke G.H., Papovich C., Cool R.J., Moustakas J., Jannuzi B.T., Kochanek C.S., et al. 2010, ApJ, 718, 1171
\bibitem[\protect\citeauthoryear{Salim et al.}{2007}]{salim07} Salim S., Rich R.M., Charlot S., Brinchmann J., Johnson B.D., Schiminovich D., Seibert M., Mallery R., et al. 2007, ApJ Suppl., 173, 267
\bibitem[\protect\citeauthoryear{Sanders et al.}{2003}]{sanders03} Sanders et al. 2003, ApJ, 126, 1607
\bibitem[\protect\citeauthoryear{Sargent et al.}{2010}]{sargent10} Sargent M.T., Schinnerer E., Murphy E., Carilli C.L., Helou G., Aussel H., Le Floc'h E., Frayer D.T., et al. 2010, ApJ, 714, L190
\bibitem[\protect\citeauthoryear{Silva et al.}{1998}]{silva98} Silva L., Granato G.L., Bressan A., Danese L., 1998, ApJ, 509, 103
\bibitem[\protect\citeauthoryear{Simpson et al.}{2014}]{simpson14} Simpson J.M., Swinbank A.M., Smail I., Alexander D.M., Brandt W.N., Bertoldi F., Chapman S.C., Coppin K.E.K., et al. 2014, ApJ, submitted (arXiv:1310.6363)
\bibitem[\protect\citeauthoryear{Solomon \& Vanden Bout}{2005}]{solomon05} Solomon P.M., Vanden Bout P.A., 2005, ARA\&A, 43, 677
\bibitem[\protect\citeauthoryear{Swinbank et al.}{2010}]{swinbank10} Swinbank A.M., Smail I., Longmore S., Harris A.I., Baker A.J., De Breuck C., Richard J., Edge A.C., Ivison R.J., Blundell R., et al. 2010, Nature, 464, 733
\bibitem[\protect\citeauthoryear{Swinbank et al.}{2014}]{swinbank14} Swinbank A.M., Simpson J.M., Smail I., Harrison C.M., Hodge J., Karim A., Walter F., Alexander D.M., et al. 2014, MNRAS, submitted (arXiv:1310.6362)
\bibitem[\protect\citeauthoryear{Takagi et al.}{2007}]{takagi07} Takagi T., Mortier A.M.J., Shimasaku K., Coppin K., Pope A., Ivison R.J., Hanami H., Serjeant S., Clements D.L., et al. 2007, MNRAS, 381, 1154
\bibitem[\protect\citeauthoryear{Taylor et al.}{2009a}]{taylor09} Taylor E.N., Franx M., van Dokkum P.G., Quadri R.F., Gawiser E., Bell E.F., Barrientos L.F., Blanc G.A., et al. 2009a, ApJ Suppl., 183, 295
\bibitem[\protect\citeauthoryear{Taylor et al.}{2009b}]{taylor09b} Taylor E.N., Franx M., van Dokkum P.G., Bell E.F., Brammer G.B., Rudnick G., Wuyts S., Gawiser E., et al. 2009b, ApJ, 694, 1171
\bibitem[\protect\citeauthoryear{Walter}{2009}]{walter09} Walter F., 2009, ASPC, 414, 287
\bibitem[\protect\citeauthoryear{Walter et al.}{2012}]{walter12} Walter F., Decarli R., Carilli C., Bertoldi F., Cox P., da Cunha E., Daddi E., Dickinson M., et al. 2012, Nature, 486, 233
\bibitem[\protect\citeauthoryear{Wardlow et al.}{2011}]{wardlow11} Wardlow J.L., Smail I., Coppin K.E.K., Alexander D.M., Brandt W.N., Danielson A.L.R., Luo B., Swinbank A.M., Walter F., Wei\ss{} A., et al. 2011, MNRAS, 415, 1479
\bibitem[\protect\citeauthoryear{Webb et al.}{2004}]{webb04} Webb T.M.A., Brodwin M., Eales S., Lilly S.J., 2004, ApJ, 605, 645
\bibitem[\protect\citeauthoryear{Wei\ss{} et al.}{2009}]{weiss09} Wei\ss{} A., Kov\'{a}cs A., Coppin K., Greve T.R., Walter F., Smail I., Dunlop J.S., Knudsen K.K., Alexander D.M., et al. 2009, ApJ, 707, 1201
\bibitem[\protect\citeauthoryear{Yun et al.}{2001}]{yun01} Yun M.S., Reddy N.A., Condon J.J., 2001, ApJ, 554, 803
\bibitem[\protect\citeauthoryear{Yun \& Carilli}{2002}]{yun02} Yun M.S., Carilli C.L., 2002, ApJ, 568, 88
\bibitem[\protect\citeauthoryear{Zibetti et al.}{2007}]{zibetti07} Zibetti S., M\'{e}nard B., Nestor D.B., Quider A.M., Rao S.M., Turnshek D.A., 2007, ApJ, 658, 161
\end{thebibliography}
\end{document}